\documentclass[aps,preprint,floatfix]{revtex4-1}
\usepackage{amsfonts}
\usepackage{amsmath}
\usepackage{amssymb}
\usepackage{dsfont}
\usepackage{hyperref}
\usepackage{color}
\usepackage{bbm} 
\usepackage{graphicx}
\usepackage{graphics}

\newcommand{\si}{\sigma}

\newcommand{\vL}{\ensuremath{\mathcal{L}}}
\newcommand{\vp}{\varphi}

\newcommand{\dslash}[1]{#1 \llap{/\kern-0.5pt}}
\newcommand{\Dslash}[1]{#1 \llap{/\kern+1.5pt}}
\newcommand{\DDslash}[1]{#1 \llap{/\kern+2.3pt}}
\newcommand{\dslashh}[1]{#1 \llap{/\kern+1pt}}

\newcommand{\bea}{\begin{eqnarray}}
\newcommand{\eea}{\end{eqnarray}}
\newcommand{\bma}{\begin{pmatrix}}
\newcommand{\ema}{\end{pmatrix}}
\newcommand{\nn}{\nonumber}

\newcommand{\DLR}{\smash{\overset{\text{\small$\leftrightarrow$}}{\smash{D}
\vphantom{+}}}}
\newcommand{\DL}{\smash{\overset{\text{\small$\leftarrow$}}{\smash{D}\vphantom{+}}}}

\begin{document}

\title{Dilepton production in the SMEFT at $\mathcal O(1/\Lambda^4)$}

\author{Radja Boughezal}
\email{rboughezal@anl.gov}
\affiliation{High  Energy  Physics  Division,  Argonne  National  Laboratory,  Argonne,  IL  60439,  USA}

\author{Emanuele Mereghetti}
\email{emereghetti@lanl.gov}
\affiliation{Theoretical  Division,  Los  Alamos  National  Laboratory,  Los  Alamos,  NM  87545,  USA}

\author{Frank Petriello}
\email{f-petriello@northwestern.edu}
\affiliation{Department  of  Physics  \&  Astronomy,  Northwestern  University,  Evanston,  IL  60208,  USA}

\begin{abstract}
We study the inclusion of $\mathcal O(1/\Lambda^4)$ effects in the Standard Model Effective Field Theory in fits to the current Drell-Yan data at the LHC. Our analysis includes the full set of dimension-6 and dimension-8 operators contributing to the dilepton process, and is performed to next-to-leading-order in the QCD coupling constant at both $\mathcal O(1/\Lambda^2)$ and $\mathcal O(1/\Lambda^4)$. We find that the inclusion of dimension-6 squared terms and certain dimension-8 operators has significant effects on fits to the current data. Neglecting them leads to bounds on dimension-6 operators off by large factors. We find that dimension-8 four-fermion operators can already be probed to the several-TeV level by LHC results, and that their inclusion significantly changes the limits found for dimension-6 operators. We discuss which dimension-8 operators should be included in fits to the LHC data. Only a manageable subset of two-derivative dimension-8 four-fermion operators need to be included at this stage given current LHC uncertainties.
\end{abstract}

\maketitle

\section{Introduction}

The Standard Model (SM) of particle physics has successfully withstood rigorous tests across a wide range of energies, from low-energy nuclear decays to high-energy  collisions. Despite its elegance and successes, the SM is not the final theory of nature, as it does not accommodate neutrino masses~\cite{Eguchi:2002dm,Ahmad:2001an}, it does not have a dark matter candidate and it cannot explain the origin of the matter-antimatter asymmetry in the universe~\cite{Gavela:1993ts,Gavela:1994ds,Gavela:1994dt,Huet:1994jb}. 
Experiments at the Large Hadron Collider (LHC) are probing the SM at the TeV scale, looking for clues that might lead to solutions to these three outstanding problems and to a better understanding of the mechanism of electroweak symmetry breaking. 
Despite a few tantalizing hints of discrepancies~\cite{LHCb:2021trn,Albahri:2021ixb,Abi:2021gix,Borsanyi:2020mff,Seng:2018yzq}, no direct evidence for new particles has so far emerged at the LHC, suggesting that the scale of new physics $\Lambda$ is larger than the electroweak scale. 

A powerful theoretical framework for investigating indirect
signatures of heavy new physics is the SM
Effective Field Theory (SMEFT). The SMEFT is formed by augmenting the SM Lagrangian with higher-dimensional operators consistent with the SM gauge symmetries and formed only from SM fields. The higher-dimensional operators in the SMEFT are suppressed by appropriate powers of a characteristic energy scale $\Lambda$ below which heavy new fields are integrated out. Complete, non-redundant bases for the dimension-6~\cite{Buchmuller:1985jz,Arzt:1994gp,Grzadkowski:2010es} and dimension-8 operators~\cite{Murphy:2020rsh,Li:2020gnx} have been constructed. Odd-dimensional operators violate lepton-number and will not be considered here.
It is an ongoing effort to analyze the numerous available data within the SMEFT framework, primarily in partial analyses of individual SMEFT sectors~\cite{Han:2004az,Cirigliano:2012ab,Chen:2013kfa,Ellis:2014dva,Wells:2014pga,Falkowski:2014tna,Cirigliano:2016nyn,deBlas:2016ojx,Hartmann:2016pil,Falkowski:2017pss,Alioli:2017ces,Alioli:2017nzr,Alioli:2018ljm,Biekotter:2018rhp,Grojean:2018dqj,Baglio:2020oqu,Boughezal:2020uwq,Boughezal:2020klp,Ricci:2020xre,Boughezal:2021kla,Ethier:2021ydt}. Recent work has been devoted to performing a global, simultaneous fit of all data available~\cite{Pomarol:2013zra,DiVita:2017eyz,Almeida:2018cld,Ellis:2018gqa,Hartland:2019bjb,Brivio:2019ius,vanBeek:2019evb,Aoude:2020dwv,Ellis:2020unq,Dawson:2020oco,Greljo:2021kvv,Ethier:2021bye},
and to study the interplay between SMEFT fits and the extraction of parton distributions from data \cite{Carrazza:2019sec,Greljo:2021kvv}.

Most of these global fits have focused on the truncation of the SMEFT expansion to dimension-6 operators at ${\cal O}(1/\Lambda^2)$. An issue that must be addressed with such an approach is the sensitivity of fits to  ${\cal O}(1/\Lambda^4)$ effects from dimension-8 operators and the square of dimension-6 terms. Intuitively their effects should be suppressed, but since many measurements at the LHC probe high energies this assumption must be tested. Furthermore, dimension-8 effects sometimes represent the leading SMEFT contributions in models with certain approximate symmetries~\cite{Liu:2016idz,Contino:2016jqw}, and their identification will be crucial to determine the UV model responsible for deviations from the SM in LHC and low-energy data \cite{Adams:2006sv,Zhang:2018shp,Zhang:2020jyn}.
Previous analyses of the impact of dimension-8 operators can be found in the literature~\cite{Degrande:2013kka,Hays:2018zze,Bellazzini:2017bkb,Bellazzini:2018paj,Ellis:2019zex,Alioli:2020kez,Murphy:2020cly,Hays:2020scx,Ellis:2020ljj}, and there is a growing body of work devoted to  study the constraints from basic principle of quantum field theory, such as unitarity and analyticity, on the allowed space of dimension-8 coefficients  \cite{Adams:2006sv,Zhang:2018shp,Zhang:2020jyn,Bi:2019phv,Remmen:2019cyz,Remmen:2020vts,Yamashita:2020gtt,Trott:2020ebl,Fuks:2020ujk}

Our goal in this work is to explore the sensitivity of high-energy Drell-Yan production of lepton pairs to dimension-8 effects in the SMEFT. Since it is calculated to high precision in the SM and is measured with residual experimental uncertainties approaching the percent level, it is an ideal channel in which to search for such effects. In previous work we have shown that a category of dimension-8 operators induce novel angular dependences not generated by QCD, and which can be potentially measured at the LHC~\cite{Alioli:2020kez}. In this work we study the impact of all sources of $1/\Lambda^4$ effects, which can arise from either genuine dimension-8 operators or from the square of dimension-6 effects, on existing experimental measurements. We include the SM next-to-next-to-leading order (NNLO) QCD and NLO electroweak corrections in the next-to-leading logarithmic approximation, as well as NLO QCD corrections to the $1/\Lambda^2$ and $1/\Lambda^4$ terms. We note that higher-order electroweak corrections to the $\mathcal (1/\Lambda^2)$ terms have recently been calculated~\cite{Dawson:2021ofa}. We summarize below the  main messages of our analysis.
\begin{itemize}

\item Effects quadratic in the dimension-6 Wilson coefficients have a significant impact on fits of the current data. For dimension-6 four-fermion operators that interfere with the SM, including $1/\Lambda^4$ effects can shift bounds on the Wilson coefficients by factors of 2-3 depending on the operator.

\item Genuine dimension-8 operators can be strongly constrained by existing Drell-Yan measurements 
at the LHC. For example, existing high-precision measurements of the Drell-Yan invariant mass distribution up to 1.5 TeV can probe dimension-8 operator scales approaching $\Lambda \approx 4$ TeV in the case of two-derivative operators of the form $\partial_\mu (\bar \psi \gamma_\nu \psi) \, \partial^\mu (\bar\chi \gamma^\nu \chi)$, with $\psi$ ($\chi$) a lepton (quark) field. 

\item Dimension-6 scalar and tensor semileptonic four-fermion operators, which contribute to the cross section at $\mathcal O(1/\Lambda^4)$, are currently probed at the same level as four-fermion operators that interfere with SM.

\item The inclusion of dimension-8 operators in the fit can significantly change the allowed regions of dimension-6 Wilson coefficients.

\end{itemize}
In the light of these results, we advocate for the inclusion of 
the dimension-6 squared contributions and of the most relevant dimension-8 operators in the analysis of LHC Drell-Yan data. At a minimum the two-derivative four-fermion operators that give contributions to the cross section scaling as ${\cal O}(s^2/\Lambda^4)$ in the high-energy limit should be included in fits to the current data to avoid misleading bounds. The number of such operators at dimension-8 is only $\mathcal O(10)$, so that the complexity of including the full dimension-8 operator set is avoided in this setup.

The paper is organized as follows. In Section \ref{basis} we provide the definition of the dimension-6 and -8 operators relevant for dilepton production. In Section \ref{CrossSection}
we sketch the calculation of the cross section and  analyze the directions in parameter space that can be probed by the Drell-Yan process. In Section \ref{Numerical}, we study the numerical impact of $\mathcal O(1/\Lambda^4)$ effects. In Sections \ref{SingleCoupling}
and \ref{Multiple} we perform a fit to the dilepton invariant mass distribution, measured at the center-of-mass energy of 8 TeV \cite{Aad:2016zzw}. To assess the sensitivity of existing data to 
$\mathcal O(1/\Lambda^4)$ effects, we first perform a single coupling analysis, in which only one operator coefficient is turned on at the new physics scale $\Lambda$. We then study to which extent dimension-8 effects can cancel dimension-6 contributions by performing a multiple parameter fit. We conclude in Section \ref{Conclusion}.

\section{Operator basis}\label{basis}

The SMEFT Lagrangian contains the most general set of operators that are invariant under the Lorentz group, the gauge group $SU(3)_c \times SU(2)_L \times U(1)_Y$, and that have the same field content as the SM, with the Higgs boson belonging to an $SU(2)_L$ doublet. SMEFT operators are organized according to their canonical dimension, with operators of higher dimension suppressed by higher powers of the new physics scale $\Lambda$.
The rapid advance of Hilbert series methods \cite{Henning:2015alf} has allowed the derivation of the complete SMEFT Lagrangian up to dimension nine \cite{Lehman:2014jma,Li:2020gnx,Murphy:2020rsh,Liao:2020jmn,Li:2020xlh}. In this work we are concerned with $\mathcal O(1/\Lambda^4)$ contributions to the Drell-Yan cross section, which arise from the square of dimension-6 operators and from the interference of dimension-8 operators with the SM.
Here we list the relevant dimension-6 and dimension-8 operators that we consider.

We first establish our notation.
The left-handed quarks and leptons and the scalar field $\varphi$ transform as doublets under $SU(2)_L$
\begin{equation}
q_L  = \left( \begin{array}{c}
u_L \\
d_L
\end{array}
\right), \qquad \ell_L = \left( \begin{array}{c}
\nu_L \\
e_L
\end{array}
\right), \qquad \varphi = \frac{v}{ \sqrt{2}} U(x) \left( 
\begin{array}{c}
0 \\
1 +  \frac{h}{v}\end{array}\right),
\end{equation}
while the right-handed quarks, $u_R$ and $d_R$, and charged leptons, $e_R$,
are singlets under $SU(2)_L$. 
$v=246$ GeV is the scalar vacuum expectation value (vev), $h$ is the physical Higgs field and $U(x)$ is a unitary matrix that encodes the Goldstone bosons.  We will denote by  $\tilde \vp$ the combination $\tilde \vp = i\tau_2 \vp^*$. 
The gauge interactions are determined by the covariant derivative
\begin{equation}
D_\mu =  \partial_\mu + i g^\prime {\rm y} B_\mu + i  \frac{g}{2} \tau^I  W^I_\mu   + i g_s G^a_\mu t^a  
\end{equation}
where $B_\mu$, $W^I_{\mu}$ and $G^a_{\mu}$ are the $U(1)_Y$, $SU(2)_L$ and $SU(3)_c$ gauge fields, respectively, and $g$, $g^\prime$, and $g_s$ are their gauge couplings.
$y$ denotes the field hypercharge assignment, given explicitly by
\begin{align}\label{hypc}
{\rm y_q}=\frac{1}{6}, \hspace{0.5cm}{\rm y_u}=\frac{2}{3}, \hspace{0.5cm}{\rm y_d}=-\frac{1}{3}, \hspace{0.5cm}{\rm y_l}=-\frac{1}{2},\hspace{0.5cm}{\rm y_e}=-1.
\end{align}
Furthermore, 
$\tau^I/2$ and $t^a$ are the $SU(2)_L$ and $SU(3)_c$ generators, in the representation of the field on which the derivative acts.

\subsection{Dimension six operators}

The dimension-6 SMEFT Lagrangian was constructed in Refs.\ \cite{Buchmuller:1985jz,Grzadkowski:2010es},   
and the operators that give the most important contributions  to Drell-Yan can be organized into three different classes:
\begin{eqnarray}\label{eq:basis1}
\mathcal L_{d6} =    \mathcal L_{\psi^2 X \varphi} +  \mathcal L_{\psi^2 \varphi^2 D}  + \mathcal L_{\psi^4}.   
\end{eqnarray}

\begin{itemize}

\item $\psi^2 X \varphi$ contains dipole couplings to the $U(1)_Y$, $SU(2)_L$ and $SU(3)_c$ gauge bosons. Here we focus on weak dipoles, which contribute at tree level:
\begin{eqnarray}
\mathcal L_{\psi^2 X \varphi} &=&  \frac{1}{\Lambda^2} \Bigg\{   \bar \ell_L \si^{\mu\nu}(C_{eB} B_{\mu\nu}+ C_{eW} {\tau}^I  {W}^I_{\mu\nu})  \varphi\, e_R +
\bar q_L \si^{\mu\nu}(C_{d B} B_{\mu\nu}+ C_{d W} {\tau}^I  {W}^I_{\mu\nu})  \varphi\, d_R \nonumber \\
& & +
\bar q_L \si^{\mu\nu}(C_{uB} B_{\mu\nu}+C_{uW} {\tau}^I  {W}^I_{\mu\nu})  \tilde{\varphi}\, u_R + 
\mathrm{h.c.}\Bigg\}\ . \label{eq:dipole}
\end{eqnarray}

\item $\psi^2 \varphi^2 D$ contains corrections to the $W$ and $Z$ boson couplings to fermions  
\begin{eqnarray}
\vL_{\psi^2 \varphi^2 D} &=& \frac{1}{\Lambda^2} \Bigg\{  \varphi^{\dagger} i \DLR_{\mu} \varphi \, \left(   \bar \ell_L   \gamma^{\mu} \, C^{(1)}_{H\ell} \ell_L +   
  \bar e_R  \gamma^{\mu}\, C_{He}  e_R\right)  +   \varphi^{\dagger}  i \DLR^I_{\mu} \varphi\,  \bar \ell_L \tau^I  \gamma^{\mu} C^{(3)}_{H\ell } \ell_L \nonumber  \\
  & & 
 +  \varphi^{\dagger} i \DLR_{\mu} \varphi \, \left(   \bar q_L   \gamma^{\mu} \, C^{(1)}_{H q } q_L +   
  \bar d_R  \gamma^{\mu}\, C_{Hd}  d_R + \bar u_R  \gamma^{\mu}\, C_{Hu}  u_R  \right)  \nonumber \\ & & +   \varphi^{\dagger}  i \DLR^I_{\mu} \varphi\,  \bar q_L \tau^I  \gamma^{\mu} C^{(3)}_{H q  } q_L  + 
\left(   \tilde{\varphi}^{\dagger}  i D_{\mu} \varphi\,  \bar u_R  \gamma^{\mu} C^{}_{H u d  } d_R + \textrm{h.c.}\right)
  \Bigg\}, \label{eq:Z}
\end{eqnarray}
where $\DLR_{\mu}=  D_\mu-\DL_\mu$, $\DLR^I_{\mu}= \tau^I D_\mu-\DL_\mu \tau^I$. The right-handed charged-current operator $C^{}_{H u d}$ contributes to $p p \rightarrow \ell \nu$, but does not induce corrections to $\ell^+ \ell^-$ production, at LO in electroweak interactions. 

\item $\mathcal L_{\psi^4}$ includes four-fermion operators. The most relevant for Drell-Yan are semileptonic four-fermion operators,
\begin{eqnarray}\label{eq:fourfermion}
\mathcal L_{\psi^4} &=& \frac{1}{\Lambda^2} \bigg\{ 
C^{(1)}_{\ell q}\, \bar \ell_L \gamma^\mu \ell_L \, \bar q_L \gamma_\mu q_L + C^{(3)}_{\ell q} \, \bar \ell_L \tau^I \gamma^\mu \ell_L \, \bar q_L  \tau^I \gamma_\mu q_L \\
& & + C_{eu} \, \bar e_R \gamma^\mu e_R \, \bar u_R \gamma_\mu u_R  +\
 C_{ed} \, \bar e_R \gamma^\mu e_R \, \bar d_R \gamma_\mu d_R \nn \\
 & & + C_{\ell u}\, \bar \ell_L \gamma^\mu \ell_L \, \bar u_R \gamma_\mu u_R +  C_{\ell d}\, \bar \ell_L \gamma^\mu \ell_L \, \bar d_R \gamma_\mu d_R  +\ C_{q e}  \, \bar e_R \gamma^\mu e_R \, \bar q_L \gamma_\mu q_L   \bigg\}  \nn \\
&& + \frac{1}{\Lambda^2}\bigg\{  C_{\ell ed q}\, \bar \ell^i_L e_R\, \bar d_R q_L^i + C^{(1)}_{\ell e q u}\, \varepsilon^{ij} \bar \ell^i_L e_R\, \bar q_L^j u_R +  C^{(3)}_{\ell e  q u}\, \varepsilon^{ij} \bar \ell^i_L \sigma^{\mu \nu} e_R\, \bar q_L^j \sigma_{\mu \nu} u_R \, + \textrm{h.c.}
  \bigg\}. \nonumber
\end{eqnarray}

\end{itemize}

Several additional operators not listed here lead to shifts of the SM couplings once the electroweak gauge boson mass matrices are diagonalized to $\mathcal O(1/\Lambda^2)$. Although we include these terms in our analysis, their impact is numerically small compared to the effects we focus on here. In the SMEFT, the EFT expansion is a double expansion in $\{v^2, s\}/\Lambda^2$, where we use the partonic center of mass energy $s$ 
to denote the typical kinematic variables in the process. As the LHC probes higher and higher scales, $s\gg v^2$, different dimension-6 and -8 operators give contributions of different importance to the cross section.    
The seven vertex corrections $C^{(1,3)}_{H\ell}$, $C^{(1,3)}_{H q}$, $C_{He}$, 
$C_{Hu}$, $C_{Hd}$ and the seven four-fermion operators  $C^{(1,3)}_{\ell q}$, $C_{eu}$, $C_{ed}$, $C_{\ell u}$, $C_{\ell d}$ and $C_{q e}$ interfere with the SM, and thus give corrections to the Drell-Yan cross section at $\mathcal O(1/\Lambda^2)$. The operators in the first class shift the value of $Z$ and $W$ boson couplings to quarks and leptons by $\mathcal O(v^2/\Lambda^2)$ with respect to the SM expectation, and thus give rise to cross sections that have the same energy behavior as the SM. 
Operators in these classes can be sensitively probed by electroweak precision data at the $Z$-pole \cite{Berthier:2015oma}. In addition, they
give important contributions to diboson production or Higgs production in association with $W$/$Z$, where they induce corrections that grow with energy \cite{Ethier:2021ydt,Aad:2020jym}.
Four-fermion operators, on the other hand, induce contributions that grow with energy and scale as $\mathcal O(s/\Lambda^2)$.
If we neglect small quark and lepton Yukawas, the dipole operators in Eq. \eqref{eq:dipole} do not interfere with the SM 
and they thus contribute to the cross section at $\mathcal O(v^2 s/\Lambda^4)$, where the power of $s$ arises from the additional derivative in the dipole interaction with respect to the SM. 
Finally, the scalar and tensor operators $C_{\ell ed q}$, $C^{(1,3)}_{\ell e q u}$
in Eq. \eqref{eq:fourfermion} contribute at $\mathcal O(s^2/\Lambda^4)$.

The Wilson coefficients of the operators in Eqs.  \eqref{eq:dipole}, 
\eqref{eq:Z} and \eqref{eq:fourfermion} are in principle matrices in flavor space. Here, for simplicity and to avoid stringent constraints from flavor physics, we choose them to be universal in both quark and lepton flavor.

\subsection{Dimension eight operators}

At dimension eight, we only consider operators that can interfere with the Standard Model. We can split the dimension-8 operators into four categories:
\begin{equation}
\mathcal L_{d8} =  \mathcal L_{\psi^4 D^2} +\mathcal L_{\psi^4 \varphi^2}+\mathcal L_{\psi^2 D^3}+\mathcal L_{\psi^2 \varphi^4 D}.
\end{equation}
As in the dimension-6 section we do not explicitly list those operators that shift the electroweak couplings at $\mathcal O(1/\Lambda^4)$.

\begin{itemize}

\item The two derivative operators are
\begin{eqnarray}\label{eq:fourfermion8}
\mathcal L_{\psi^4 D^2} &=& \frac{1}{\Lambda^4} \bigg\{ 
C^{(1)}_{\ell^2\, q^2\,  D^2}\, \partial_\nu \left( \bar \ell_L \gamma^\mu \ell_L\right) \, \partial^\nu \left(\bar q_L \gamma_\mu q_L\right) + 
C^{(3)}_{\ell^2\, q^2\,  D^2}\, D_\nu \left( \bar \ell_L \gamma^\mu \tau^I \ell_L\right) \, D^\nu \left(\bar q_L \gamma_\mu \tau^I q_L\right) \nonumber
\\
& & + C^{(1)}_{e^2\,u^2\, D^2} \, \partial_\nu \left(\bar e_R \gamma^\mu e_R \right)\, \partial^\nu \left( \bar u_R \gamma_\mu u_R \right) +\
 C^{(1)}_{e^2\,d^2\, D^2} \, \partial_\nu \left( \bar e_R \gamma^\mu e_R \right) \, \partial^\nu \left( \bar d_R \gamma_\mu d_R \right) \nn \\
 & & + C^{(1)}_{ \ell^2\,u^2\, D^2}\, \partial_\nu \left( \bar \ell_L \gamma^\mu \ell_L \right) \, \partial^\nu \left( \bar u_R \gamma_\mu u_R\right) +  C^{(1)}_{\ell^2\, d^2\, D^2}\, \partial_\nu \left( \bar \ell_L \gamma^\mu \ell_L\right) \, \partial^\nu \left( \bar d_R \gamma_\mu d_R\right) \nonumber \\ & & +\ C^{(1)}_{ q^2\, e^2\, D^2}  \, \partial_\nu \left( \bar e_R \gamma^\mu e_R\right) \, \partial^\nu \left( \bar q_L \gamma_\mu q_L \right)  \bigg\}.  
\end{eqnarray}
These operators interfere with the SM  to generate a $\mathcal O(s^2/\Lambda^4)$ correction to the cross section.
The other class of two-derivative operators has the form\footnote{In Ref. \cite{Alioli:2020kez}, we adopted a definition of 
$C^{(2)}_{\ell^2 q^2 D^2}$ and of the analogous operators in other helicity channels 
without the symmetrization over $\mu$, $\nu$. With respect to the definition in Ref. \cite{Alioli:2020kez},  Eq. \eqref{C2} has the advantages that the interference of 
$C^{(2)}_{\ell^2 q^2 D^2}$ with the SM vanishes in the dilepton invariant mass distribution, and that  $C^{(2)}_{\ell^2 q^2 D^2}$ does not mix with $C^{(1)}_{\ell^2 q^2 D^2}$ at $\mathcal O(\alpha_s)$ in QCD.}
\begin{eqnarray}\label{C2}
C^{(2)}_{\ell^2 q^2 D^2}  \, \bar q_L  \gamma^{(\mu} \overleftrightarrow D^{\nu )} q_L \, \bar{\ell}_L \gamma_{(\mu} \overleftrightarrow{D}_{\nu)} \ell_L,
\label{eq:class2FF}
\end{eqnarray}
where the notation $(\mu \nu)$ denotes symmetrization over the indices $\mu$ and $\nu$
\begin{equation}
 \gamma^{(\mu} \overleftrightarrow D^{\nu )} =  \left( \gamma^{\mu} \overleftrightarrow D^{\nu } +
 \gamma^{\nu} \overleftrightarrow D^{\mu }\right).
\end{equation}
These operators give rise to interesting angular distributions, 
which we considered in Ref. \cite{Alioli:2020kez}. The interference with the SM, however, vanishes once we integrate over the lepton angle $\cos\theta$.
While the cuts on the leptons transverse momenta and rapidities 
prevent an exact cancellation, these operators cannot be efficiently probed with the distributions we study in this paper.  
 
\item There are several semileptonic operators with two Higgses:
 \begin{eqnarray}\label{eq:qqllphi}
\mathcal L_{\psi^4 \varphi^2} &=& \frac{1}{\Lambda^4} \Bigg\{  C^{(1)}_{\ell^2 q^2 H^2} \bar q_L \gamma^\mu q_L \, \bar{\ell}_L \gamma_\mu \ell_L \, \varphi^{\dagger} \varphi + 
C^{(3)}_{\ell^2 q^2 H^2} \bar q_L \tau^I \gamma^\mu q_L \, \bar{\ell}_L \tau^I \gamma_\mu \ell_L \, \varphi^{\dagger} \varphi  \nn \\
& &+ C^{(4)}_{\ell^2 q^2 H^2} \bar q_L \tau^I \gamma^\mu q_L \, \bar{\ell}_L \gamma_\mu \ell_L \, \varphi^{\dagger}\tau^I  \varphi +
C^{(2)}_{\ell^2 q^2 H^2} \bar q_L  \gamma^\mu q_L \, \bar{\ell}_L \tau^I \gamma_\mu \ell_L \, \varphi^{\dagger}\tau^I  \varphi \nn \\ & &+
C^{(5)}_{\ell^2 q^2 H^2} \varepsilon^{IJK} \bar q_L  \tau^I \gamma^\mu q_L \, \bar{\ell}_L \tau^J \gamma_\mu \ell_L \, \varphi^{\dagger}\tau^K  \varphi \nn
\\
& & + C^{(1)}_{e^2 u^2 H^2} \bar u_R \gamma^\mu u_R \, \bar{e}_R \gamma_\mu e_R\varphi^\dagger \varphi
+  C^{(1)}_{e^2 d^2 H^2} \bar d_R \gamma^\mu d_R \, \bar{e}_R \gamma_\mu e_R \varphi^\dagger \varphi \nn \\
& & +  C^{(1)}_{\ell^2 u^2 H^2} \bar u_R \gamma^\mu u_R \, \bar{\ell}_L \gamma_\mu \ell_L \varphi^\dagger \varphi 
+  C^{(2)}_{\ell^2 u^2 H^2} \bar u_R \gamma^\mu u_R \,\bar{\ell}_L \gamma_\mu \tau^J  \ell_L \varphi^\dagger \tau^J \varphi  \nonumber \\
& & +  C^{(1)}_{\ell^2 d^2 H^2} \bar d_R \gamma^\mu d_R \, \bar{\ell}_L \gamma_\mu \ell_L \varphi^\dagger \varphi 
+  C^{(2)}_{\ell^2 d^2 H^2} \bar d_R \gamma^\mu d_R \,\bar{\ell}_L \gamma_\mu \tau^J  \ell_L \varphi^\dagger \tau^J \varphi  \nonumber \\
& & +   C^{(1)}_{q^2 e^2 H^2} \bar q_L \gamma^\mu q_L \, \bar{e}_R \gamma_\mu e_R \varphi^\dagger \varphi 
+  C^{(2)}_{q^2 e^2 H^2} \bar q_L \gamma^\mu \tau^J q_L \,\bar{e}_R \gamma_\mu  e_R \varphi^\dagger \tau^J \varphi \Bigg\}.
\end{eqnarray}
$C^{(5)}_{\ell^2 q^2 H^2}$ does not interfere with the SM. The net effect of the other operators is to provide an independent coefficient in the 12 channels $e_{i} u_{j}$, $e_{i} d_j$, $\nu_L u_j$, $\nu_L d_j$ with $i, j \in \{ L, R\}$. The contributions
of the operators in Eq. \eqref{eq:qqllphi} to each flavor and helicity channel are given in Eq. \eqref{Vff2}.

\item The next class we consider are fermion bilinear operators with three derivatives. They give rise to vertices of the form $\bar f \gamma^\mu f \partial^2 Z_\mu$:
\begin{eqnarray}\label{qqphiphid3}
& & \mathcal L_{\psi^2 D^3} = \frac{1}{\Lambda^4} \Bigg\{ C^{(1)}_{\ell^2 H^2 D^3}\,  i \bar \ell_L \gamma^\mu D^\nu \ell_L \, (D_{(\mu} D_{\nu)} \varphi)^{\dagger} \,  \varphi
+ C^{(2)}_{\ell^2 H^2 D^3}\,  i \bar \ell_L \gamma^\mu D^\nu \ell_L \,  \varphi^{\dagger} \, D_{(\mu} D_{\nu)} \varphi \nonumber \\
& &+ C^{(3)}_{\ell^2 H^2 D^3}\,  i \bar \ell_L \gamma^\mu \tau^I D^\nu \ell_L \, (D_{(\mu} D_{\nu)} \varphi)^{\dagger} \tau^I \,  \varphi
+ C^{(4)}_{\ell^2 H^2 D^3}\,  i \bar \ell_L \gamma^\mu \tau^I D^\nu \ell_L \,  \varphi^{\dagger} \tau^I \, D_{(\mu} D_{\nu)} \varphi 
\Bigg\} \nonumber \\
& &+ \frac{1}{\Lambda^4} \Bigg\{ C^{(1)}_{e^2 H^2 D^3}\,  i \bar e_R \gamma^\mu D^\nu e_R \, (D_{(\mu} D_{\nu)} \varphi)^{\dagger} \,  \varphi
+ C^{(2)}_{e^2 H^2 D^3}\,  i \bar e_R \gamma^\mu D^\nu e_R \,  \varphi^{\dagger} \, D_{(\mu} D_{\nu)} \varphi \Bigg\} \nonumber \\
& &+ \frac{1}{\Lambda^4} \Bigg\{ C^{(1)}_{q^2 H^2 D^3}\,  i \bar q_L \gamma^\mu D^\nu q_L \, (D_{(\mu} D_{\nu)} \varphi)^{\dagger} \,  \varphi
+ C^{(2)}_{q^2 H^2 D^3}\,  i \bar q_L \gamma^\mu D^\nu q_L \,  \varphi^{\dagger} \, D_{(\mu} D_{\nu)} \varphi \nonumber \\
& &+ C^{(3)}_{q^2 H^2 D^3}\,  i \bar q_L \gamma^\mu \tau^I D^\nu q_L \, (D_{(\mu} D_{\nu)} \varphi)^{\dagger} \tau^I \,  \varphi
+ C^{(4)}_{q^2 H^2 D^3}\,  i \bar q_L \gamma^\mu \tau^I D^\nu q_L \,  \varphi^{\dagger} \tau^I \, D_{(\mu} D_{\nu)} \varphi 
\Bigg\} \nonumber \\
& &+ \frac{1}{\Lambda^4} \Bigg\{ C^{(1)}_{u^2 H^2 D^3}\,  i \bar u_R \gamma^\mu D^\nu u_R \, (D_{(\mu} D_{\nu)} \varphi)^{\dagger} \,  \varphi
+ C^{(2)}_{u^2 H^2 D^3}\,  i \bar u_R \gamma^\mu D^\nu u_R \,  \varphi^{\dagger} \, D_{(\mu} D_{\nu)} \varphi \Bigg\} \nonumber \\
& &+ \frac{1}{\Lambda^4} \Bigg\{ C^{(1)}_{d^2 H^2 D^3}\,  i \bar d_R \gamma^\mu D^\nu d_R \, (D_{(\mu} D_{\nu)} \varphi)^{\dagger} \,  \varphi
+ C^{(2)}_{d^2 H^2 D^3}\,  i \bar d_R \gamma^\mu D^\nu d_R \,  \varphi^{\dagger} \, D_{(\mu} D_{\nu)} \varphi \Bigg\}  \nonumber \\ & & + \textrm{h.c.}.
\end{eqnarray}
Only six linear combinations contribute to Drell-Yan. 
We  introduce the coupling to left-handed electrons
\begin{eqnarray}\label{Cel2H2D3}
 C^{(e)}_{\ell^2 H^2 D^3} = C^{(1)}_{\ell^2 H^2 D^3} -  C^{(2)}_{\ell^2 H^2 D^3} + 
 C^{(3)}_{\ell^2 H^2 D^3}  - C^{(4)}_{\ell^2 H^2 D^3}. 
\end{eqnarray}
The other linear combinations are $C^{(1)}_{f^2_R H^2 D^3} - C^{(2)}_{f^2_R H^2 D^3}$,
with $f_R \in  \{e, u, d\}$, and 
$C^{(1)}_{q^2 H^2 D^3} - C^{(2)}_{q^2 H^2 D^3}$, $C^{(3)}_{q^2 H^2 D^3} - C^{(4)}_{q^2 H^2 D^3}$.

\item The final corrections we consider are fermion bilinear operators with a single derivative. These give rise to momentum independent corrections to the $Z$-boson vertices. We can write the relevant operators as
\begin{eqnarray}
& &\mathcal L_{\psi^2 \varphi^4 D} = \frac{1}{\Lambda^4} \Bigg\{
C_{e^2H^4D} (\bar e_R \gamma^\mu e_R) (\varphi^{\dagger} \overleftrightarrow D_{\mu} \varphi )(\varphi^{\dagger} \varphi) + 
C^{(1)}_{\ell^2H^4D} \, i (\bar \ell_L \gamma^\mu \ell_L) (\varphi^{\dagger} \overleftrightarrow D_{\mu} \varphi )(\varphi^{\dagger} \varphi) \nonumber \\ & &   +C^{(2)}_{\ell^2H^4D} \, i (\bar \ell_L \gamma^\mu \tau^I \ell_L) \left[ (\varphi^{\dagger} \overleftrightarrow D^I_{\mu} \varphi )(\varphi^{\dagger} \varphi)+ (\varphi^{\dagger} \overleftrightarrow D_{\mu} \varphi )(\varphi^{\dagger} \tau^I \varphi) \right] \nonumber \\ 
&&   + C^{(1)}_{q^2H^4D} \, i (\bar q_L \gamma^\mu q_L) (\varphi^{\dagger} \overleftrightarrow D_{\mu} \varphi )(\varphi^{\dagger} \varphi)  \nonumber \\ 
&& + C^{(2)}_{q^2H^4D} \, i (\bar q_L \gamma^\mu \tau^I q_L) \left[ (\varphi^{\dagger} \overleftrightarrow D^I_{\mu} \varphi )(\varphi^{\dagger} \varphi)+(\varphi^{\dagger} \overleftrightarrow D_{\mu} \varphi )(\varphi^{\dagger}  \tau^I \varphi) \right]\nonumber\\
&& +C_{u^2H^4D} (\bar u_R \gamma^\mu u_R) (\varphi^{\dagger} \overleftrightarrow D_{\mu} \varphi )(\varphi^{\dagger} \varphi) 
+ C_{d^2H^4D} (\bar d_R \gamma^\mu d_R) (\varphi^{\dagger} \overleftrightarrow D_{\mu} \varphi )(\varphi^{\dagger} \varphi) 
 \bigg\}
\label{eq:f2H4D}
\end{eqnarray}
This class of operators introduces seven independent Wilson coefficients. Since they give rise only to momentum-independent vertex corrections they are difficult to disentangle from similar effects at dimension-6, and are typically much smaller than other shifts of the cross section induced by SMEFT operators.

\end{itemize}

The corrections to the cross section from the operators in Eqs. \eqref{eq:qqllphi} and \eqref{qqphiphid3} 
scale as $\mathcal O(v^2 s/\Lambda^4)$, while those in Eq.~(\ref{eq:f2H4D}) scale as  $\mathcal O(v^4/\Lambda^4)$.
For dimension-8 operators we also make the assumption of flavor universality.

\section{Calculation of the cross section}\label{CrossSection}

The Drell-Yan cross section has the structure 
\begin{equation}\label{sigmascheme}
 \frac{d\sigma}{ d m_{\ell \ell}} = \frac{d\sigma_{\rm SM}}{ d m_{\ell \ell}}
 + \sum_{i } \left(\frac{ a^{(6)}_i(m_{\ell \ell})}{\Lambda^2} \, C^{(6)}_i +
\frac{ a^{(8)}_i(m_{\ell \ell})}{\Lambda^4} C^{(8)}_i \right)  
 + \sum_{i j} \frac{b^{(6)}_{ij}(m_{\ell \ell})}{\Lambda^4} \, C^{(6)}_i C^{(6)}_j,
 \end{equation}
where we collectively denote by $C^{(6)}$ and $C^{(8)}$ the coefficients of dimension-6 and dimension-8 operators, respectively. 
As we already mentioned, the chiral structure of the SM implies that only a limited number of interference terms exist. Similarly, if we neglect small lepton and quark Yukawas, the interference terms  between different dimension-6 operators, $b^{(6)}_{ij}$, are limited to interference between the purely left-handed operators $C_{\ell q}^{(1)}$ and $C_{\ell q}^{(3)}$, between the  $SU(2)_L$ and $U(1)_Y$ dipole operators,
and between scalar and tensor operators. The latter vanishes when integrating over the 
angular variables, leaving some small residual effects due to the cuts on the lepton
transverse momenta and rapidities.  

A detailed discussion of the SMEFT vertices that enter the Drell-Yan cross section is given in Appendix~\ref{sec:Vparam}. We summarize here the main points of this discussion. At the $\mathcal O(1/\Lambda^4)$ level, three distinct contributions to the Drell-Yan cross section are possible.
\begin{itemize}

\item The most important class of corrections scales as $\mathcal O(s^2/\Lambda^4)$. Examples of operators that lead to this dependence are momentum-dependent four-fermion operators at dimension-8, and scalar/tensor dimension-6 four-fermion operators. These terms have a large effect on the cross section, and our numerical fits in the next section show that they cannot be neglected in fits to the current data.

\item The second type of correction scales as $\mathcal O(v^2 s/\Lambda^4)$. Dimension-6 dipole operators and momentum-dependent $Z$-vertex corrections at dimension-8 lead to this behavior. The impact of the these terms on fits to the Drell-Yan data is smaller than those in the previous category.

\item Finally, the last type of correction scales as $\mathcal O(v^4/\Lambda^4)$. These terms come from momentum-independent $Z$-boson vertex corrections at dimension-8. Their effects on the Drell-Yan cross section are small for all reasonable choices of parameters.

\end{itemize}
An enumeration of the Wilson coefficients that appear in the Drell-Yan cross section at each order in the $\mathcal O(1/\Lambda)$ expansion is given in the Appendix. The number of Wilson coefficients that enters the Drell-Yan cross section does not increase dramatically upon going from dimension-6 to dimension-8, increasing from approximately 30 to 60. One reason for this is that the dimension-8 scalar, tensor and dipole operators do not interfere with the SM amplitude and therefore have no effect at $\mathcal O(1/\Lambda^4)$. Such operators at dimension-6 can interfere with themselves and contribute at $\mathcal O(1/\Lambda^4)$. We also note that not all of these Wilson coefficients can be independently measured in the Drell-Yan process. The simplest way to see this is to count the number of vertex structures that appear at each order in $\mathcal O(1/\Lambda)$. The number of such structures sets the upper limit on how many combinations of parameters are in principle distinguishable at each order. We note that not every term may be distinguishable in practice given limited experimental measurements, or approximate degeneracies between the behavior of different vertices. We also note that for some structures, such as the momentum-dependent four-fermion vertices defined in Eq.~(\ref{eq:V8def}) that first appear at $\mathcal O(1/\Lambda^4)$, the number of Wilson coefficients is greater than the number of independent interaction structures. The vertices are given in the Appendix, and include both four-fermion interactions and corrections to gauge boson-fermion vertices. Structures with different momentum dependence can be distinguished through measurements of the invariant mass distributions, 
up-quark and down-quark structures can in principle be distinguished through their different rapidity dependence,
while angular distributions can disentangle vector from dipole, scalar and tensor operators.
A total of 14 structures contributes at $\mathcal O(1/\Lambda^2)$. An additional 32 enter at $\mathcal O(1/\Lambda^4)$, including the two-derivative operators discussed in Eq.~(\ref{eq:class2FF}) but not explicitly considered here.

We compute $a^{(6)}$, $a^{(8)}$ and $b^{(6)}$ at NLO in QCD.
A first $\mathcal O(\alpha_s)$ effect arises from the renormalization
group evolution of the Wilson coefficients of SMEFT operators. 
The dimension-6 corrections to the $Z$ and $W$ boson couplings,
the vector-like semileptonic operators and the dimension-8 operators 
in Eqs. \eqref{eq:fourfermion8}, \eqref{eq:qqllphi}, \eqref{qqphiphid3}
and \eqref{eq:f2H4D} do not run at one loop in QCD. 
For the dipole, scalar and tensor operators, we evolve the coefficients from the scale $\mu_0$, chosen to be close to the new physics scale $\Lambda$, to the renormalization scale $\mu_R$, using two-loop anomalous dimensions \cite{Misiak:1994zw,Vermaseren:1997fq,Degrassi:2005zd}; see Appendix \ref{RGE} and Ref. \cite{Alioli:2018ljm} for more details. 
The renormalization of the two-derivatives operator  $C^{(2)}_{\ell^2 q^2 D^2}$, and similar operators with different quark and lepton chiralities, has been studied in the context of higher-twist operators, and it is known to three-loops  \cite{Gracey:2003mr}.
The second effect arises from QCD virtual and real emissions. 
For the dimension-6 and dimension-8 operators with the same chiral structure as the SM, these corrections are identical to QCD corrections to SM amplitudes.  For dipole, scalar, and tensor operators we use the calculation of Ref. \cite{Alioli:2018ljm}, 
while we compute the corrections induced by $C^{(2)}_{\ell^2 q^2 D^2}$.
Our calculation is complete at $\mathcal O(\alpha_s/\Lambda^2)$.
At $\mathcal O(\alpha_s/\Lambda^4)$, we include all corrections 
proportional to operators that contribute at the Born level, and that thus generate contributions enhanced in the soft and collinear limits. We do not include corrections from dimension-8 operators with gluons, such as  
\begin{eqnarray}\label{Cl2q2G}
C^{(1)}_{\ell^2 q^2 G}  \, \bar q_L  t^a  \gamma^{\mu } q_L \, \bar{\ell}_L \gamma^{\nu}  \ell_L G^{a}_{\mu \nu}.
\end{eqnarray}
While these operators induce small corrections to the dilepton invariant mass distribution, 
and we can safely neglect them, they might play a more prominent role in studies of the dilepton transverse momentum distribution. 

In the invariant mass bins we consider, NLO QCD corrections increase the SM cross section by about 20\%, reaching 26\% in the highest bin. SMEFT cross sections receive contributions of similar size, about $30\%$ for scalar, tensor and dipole operators.

\section{Numerical impact of dimension-6 and dimension-8 operators}
\label{Numerical}

\begin{figure}
 \includegraphics[width=\columnwidth]{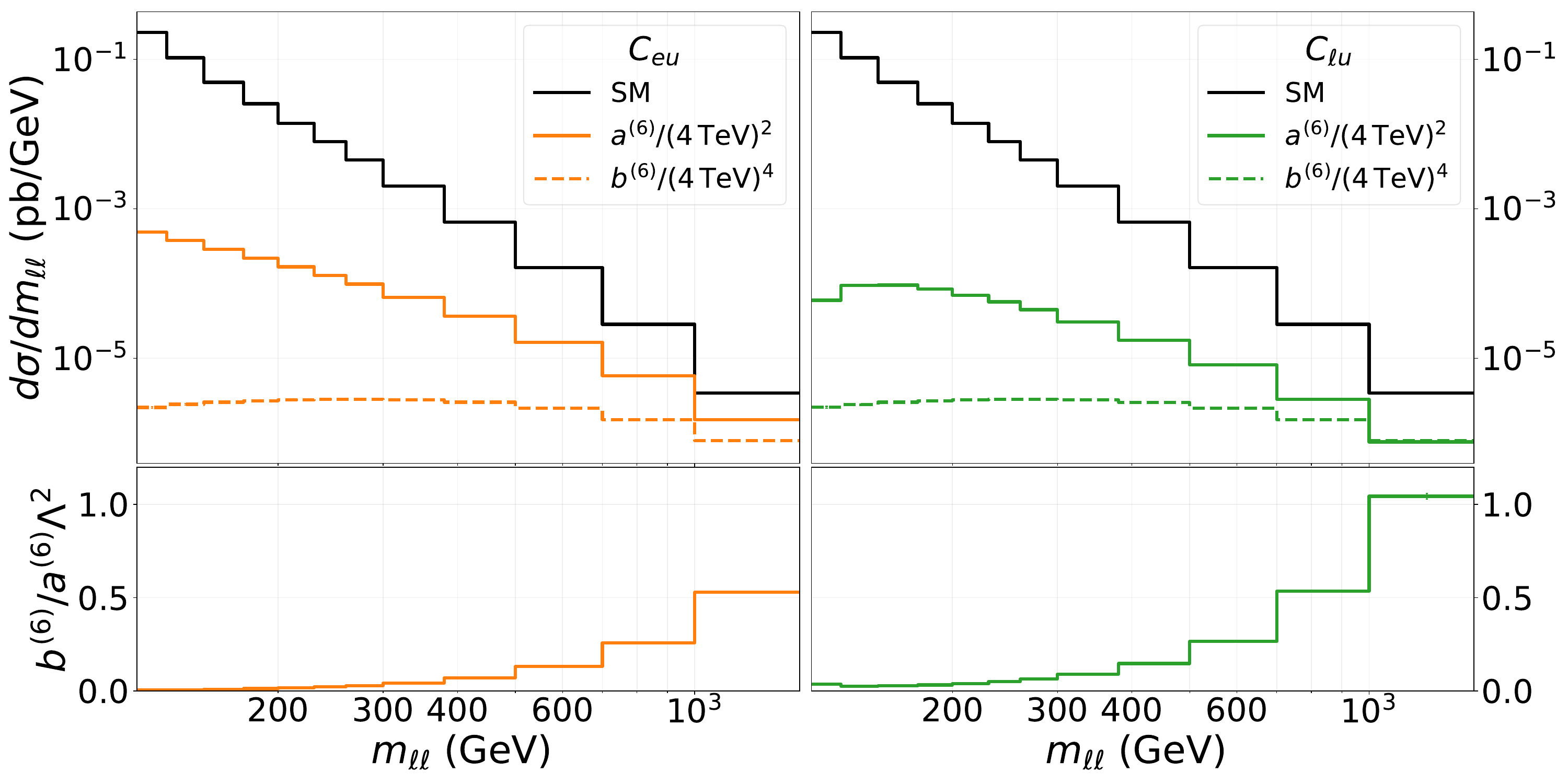}
 \caption{$a^{(6)}$ and $b^{(6)}$ coefficients for the operators $C_{eu}$ and $C_{\ell u}$, for $\Lambda= 4$ TeV.\label{Figure1}}
\end{figure}

Before discussing the constraints and the impact of $\mathcal O(\Lambda^{-4})$ corrections, in Figure~\ref{Figure1}  we show the linear ($a^{(6)}$) and quadratic ($b^{(6)}$)
corrections to the SM cross sections for the representative SMEFT coefficients $C_{eu}$ and $C_{\ell u}$, for the choice of UV scale $\Lambda =4$ TeV. For this choice of UV scale we have $(s/\Lambda)^2 \ll 1$ since the maximum invariant mass probed is 1.5~TeV, and we expect the EFT expansion to be well behaved. 
In the case of $C_{eu}$, we see that for $\Lambda = 4$ TeV the quadratic term is negligible up to about $m_{\ell \ell} \sim 500$ GeV, while it becomes approximately 25\% and  50\% of the linear piece in the two highest invariant mass bins. The quadratic and linear terms become comparable in the highest invariant mass bin, $m_{\ell \ell} \in \left[1.0,1.5\right]$ TeV, for $\Lambda=3$ TeV, which we also expect to be in the range of validity of the EFT.
For $C_{\ell u}$, $b^{(6)}$ is even more important, being, in the highest bin, larger than the linear term at 
$\Lambda = 4$ TeV and  70\% of the linear term at $\Lambda = 5$ TeV. 
The values of the coefficients $a^{(6)}/\Lambda^2$ and $b^{(6)}/\Lambda^4$
induced by the dimension-6 SMEFT operators that can interfere with the SM are given in Tables \ref{Vector6a}
and \ref{Vector6b} in  Appendix \ref{CStables} for $\Lambda = 4$ TeV. In the highest invariant mass bin, $m_{\ell \ell} \in \left[1.0,1.5\right]$ TeV, the quadratic term ranges from half to two times $a^{(6)}$. We thus expect quadratic contributions to have a significant impact on the coefficient fits.

\begin{figure}
 \includegraphics[width=\textwidth]{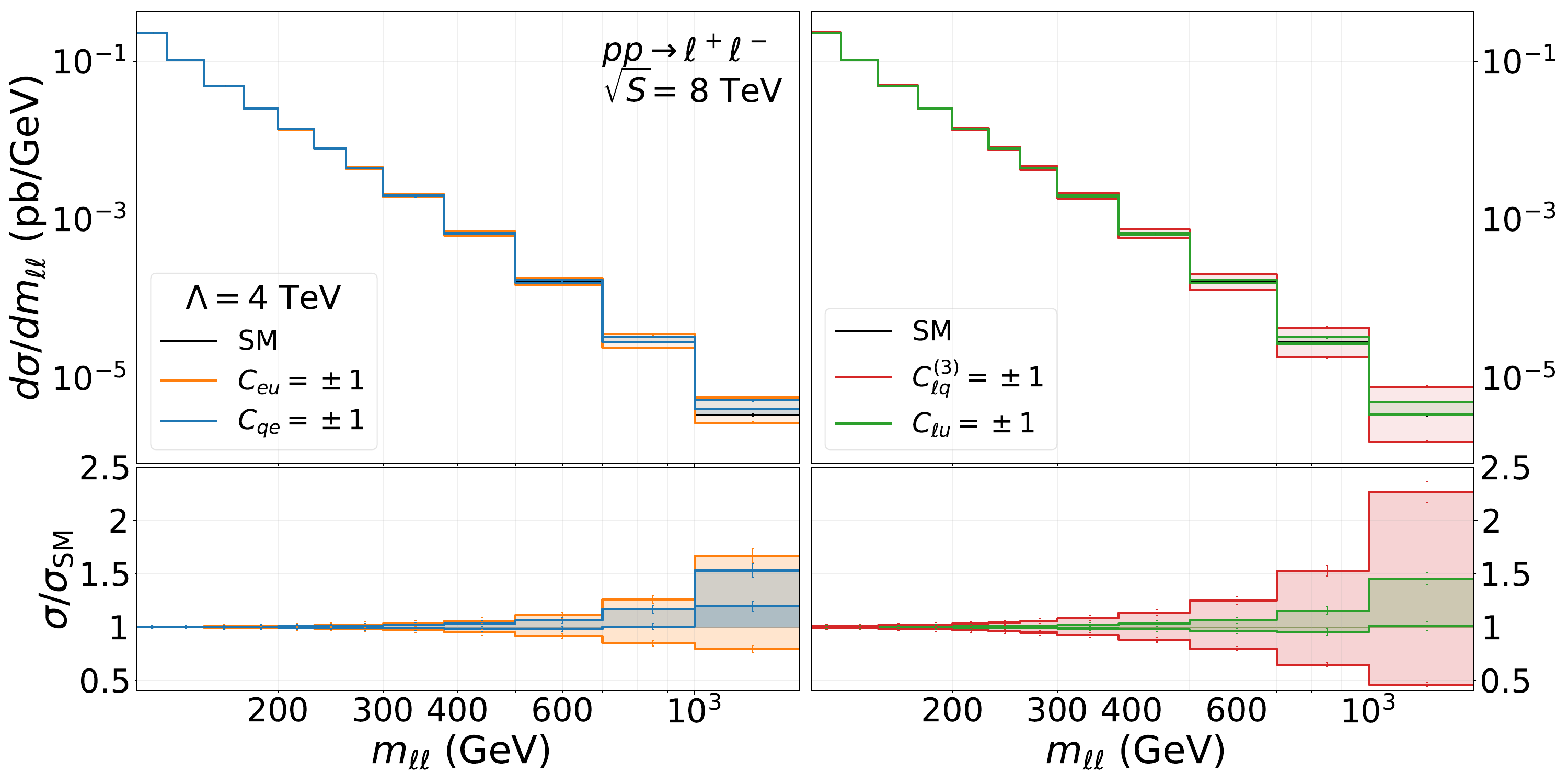}
 \caption{Contributions of vector and axial operators to dilepton production. The colored bands denote the deviation from the Standard Model cross sections, as the couplings vary between $\pm 1$ for $\Lambda = 4$ TeV.}
 \label{Fig_mll_clq}
\end{figure}

\begin{figure}
 \includegraphics[width=\textwidth]{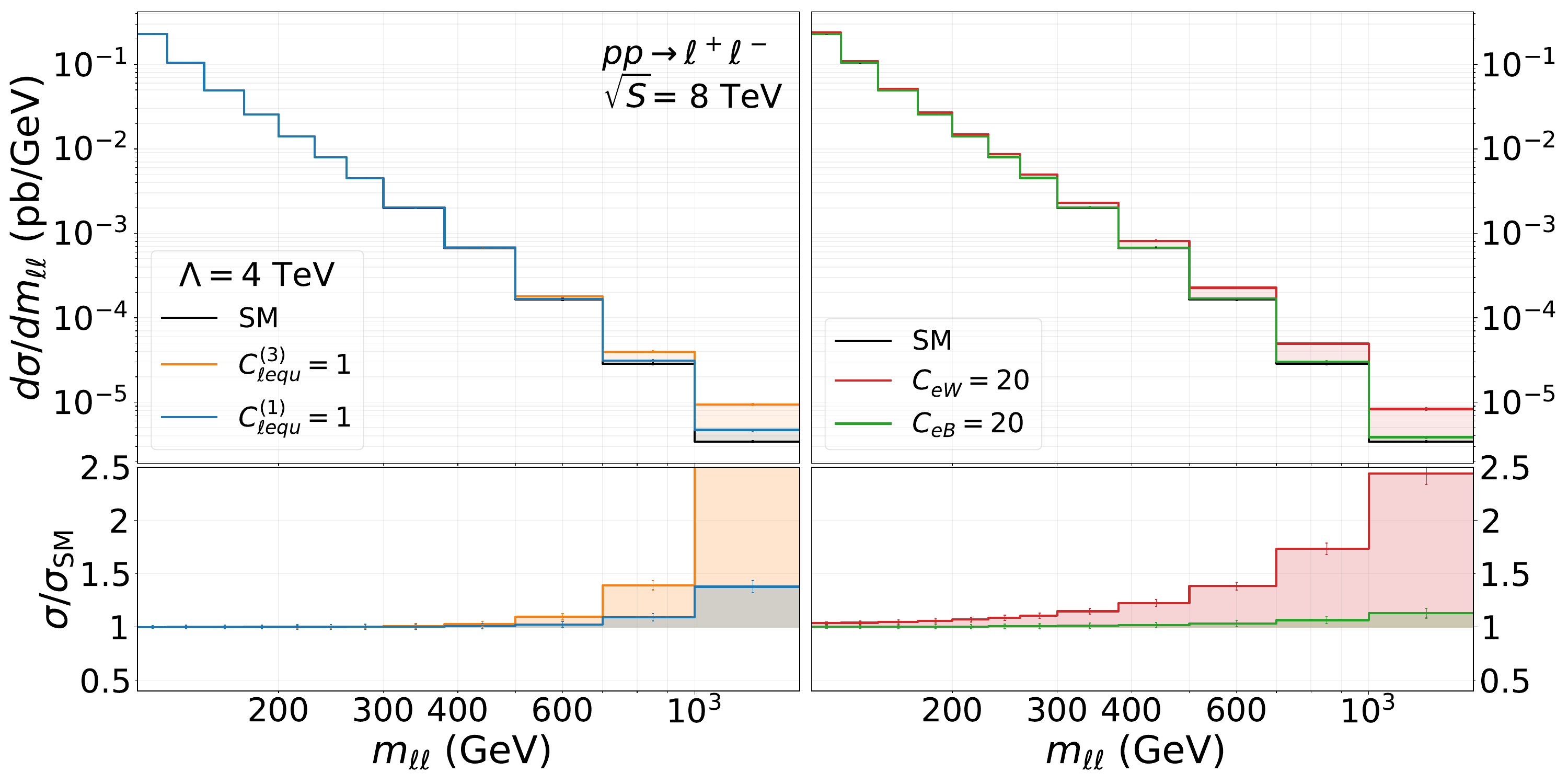}
 \caption{Contributions of scalar, tensor and dipole operators to dilepton production.
 The colored bands denote the deviation from the Standard Model cross sections, as the couplings vary between 0 and 1 for $C^{(1,3)}_{\ell e q u}(\mu_0)$
 and between $0$ and $20$ for $C^{}_{eW,\, eB}(\mu_0)$, with $\Lambda = 4$ TeV. The Wilson coefficients are defined at the initial scale  $\mu_0 =1$ TeV. 
 }
 \label{Fig_mll_clequ}
\end{figure}

\begin{figure}
 \includegraphics[width=\textwidth]{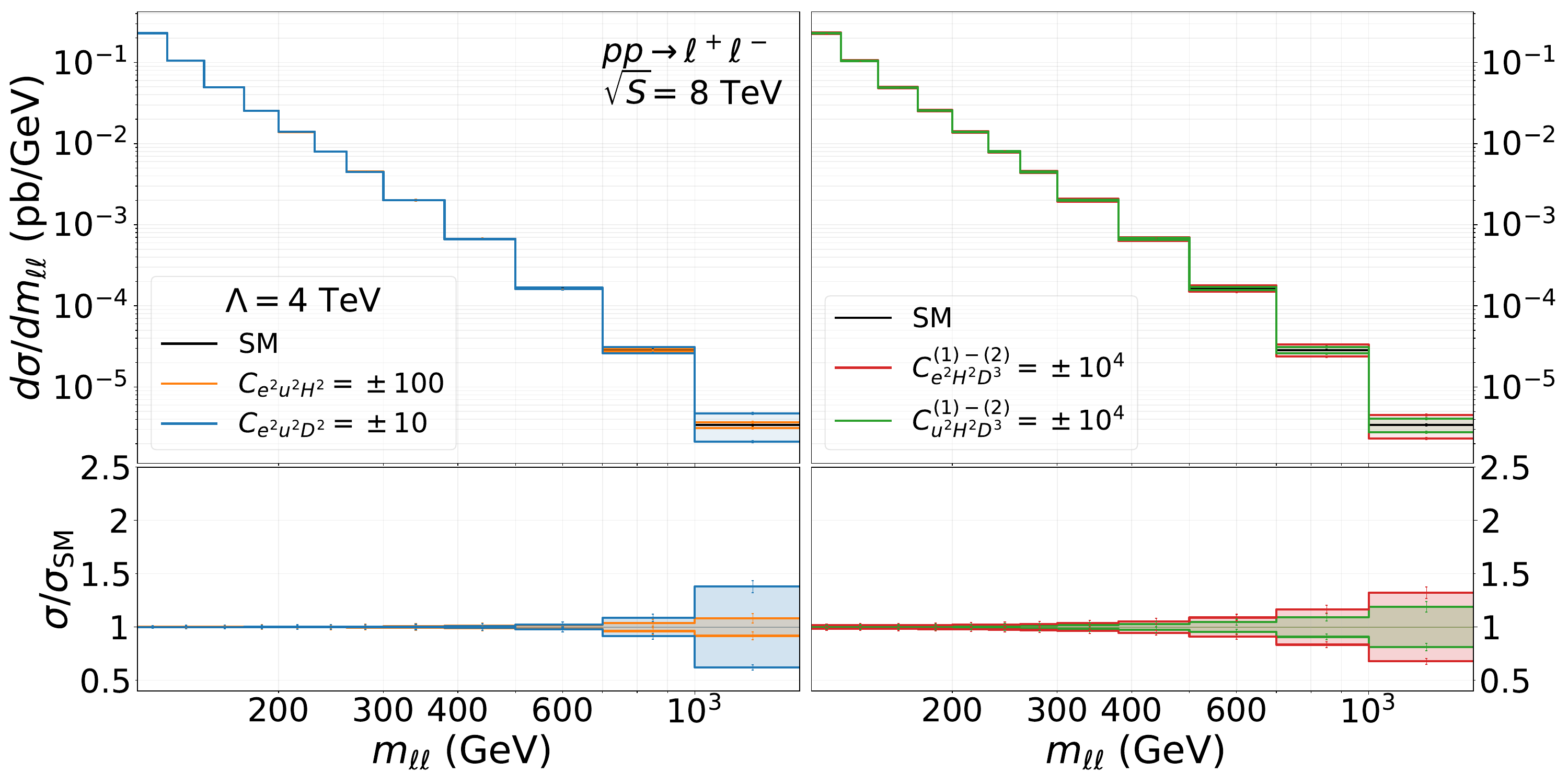}
 \caption{Contributions of dimension-8 operators to dilepton production.
 The colored bands denote the deviation from the Standard Model cross sections, as the couplings vary between $\pm 10$ for $C^{}_{e^2 u^2 D^2}$, 
 $\pm 100$ for $C^{}_{e^2 u^2 H^2}$
and $\pm 10^4$ for $C^{}_{e^2 H^2 D^3}$ and $C^{}_{u^2 H^2 D^3}$, with $\Lambda = 4$ TeV.
 }
 \label{Fig_mll_eight}
\end{figure}

In Figures~\ref{Fig_mll_clq}, \ref{Fig_mll_clequ}, and~\ref{Fig_mll_eight} we examine the behavior of vector-like four-fermion operators, scalar, tensor and dipole operators and dimension-8 operators, respectively.  All vector operators induce comparable corrections to the SM cross sections, with some enhancement for the operators that couple to $u$ quarks. The size of the interference with the SM is dictated by the SM $Z$ boson and photon couplings. Scalar and tensor four-fermion operators do not interfere with the SM. However, the left panel of Fig. \ref{Fig_mll_clequ} shows that these operators induce large corrections to the Drell-Yan cross section, comparable to (if not larger than) vector operators. Since these operators
always increase the SM cross section and do not leave room for cancellations, we will see that, in a single coupling analysis, the bounds on scalar and tensor coefficients are stronger than on vector operators. 
The dipole operators also do not interfere with the SM. In their case, however, the correction to the SM cross section grows as $s/\Lambda^2$
compared to $s^2/\Lambda^4$ for four-fermion operators.
For $\Lambda =4$ TeV, the dimensionless dipole coefficients $C_{f W}$
and $C_{f B}$ need to be larger than four-fermion coefficients by about a factor of ten to cause comparable corrections to the cross section. 
The corrections to $d\sigma/d m_{\ell \ell}$ induced by dipole, scalar and tensor operators, in the binning of Ref. \cite{Aad:2016zzw}, are given in Tables \ref{DipoleU}, \ref{DipoleD} and \ref{Scalar}.

Finally in Fig. \ref{Fig_mll_eight} we show the corrections induced by 
four dimension-8 operators that couple right-handed quarks and electrons: the two-derivative operator $C_{e^2 u^2 D^2}$,
the correction to the dimension-6 coupling $C_{e^2 u^2 H^2}$,
and two derivative couplings of the $Z$ boson to quark and leptons,
$C_{e^2 H^2 D^3}$ and $C_{u^2 H^2 D^3}$. We see that the derivative operator induces sizable corrections to the cross section. 
At $\Lambda = 4$ TeV, a dimensionless coupling of order 10, corresponding to a UV scale of 2 TeV, gives a 40 \%
corrections to the Drell-Yan cross section in the invariant mass bin 
$m_{\ell \ell} \in [1.0,1.5]$ TeV.
The corrections are smaller in the case of  $C_{e^2 u^2 H^2}$. Its contribution to the cross section becomes visible for  $C_{e^2 u^2 H^2} = \mathcal O(100)$, corresponding to an effective scale of about 1 TeV. The $Z$-boson form factor operators need very low scales, $\Lambda \sim 400$ GeV, assuming Wilson coefficients of ${\cal O}(1)$, to induce large corrections. For such a low scale, of course, a SMEFT analysis of the data in Ref. \cite{Aad:2016zzw} is not justified. We note that this conclusion is valid only when assuming that the underlying UV completion giving rise to the SMEFT is weakly coupled. The dimensionless Wilson coefficients generically behave as $C_i \sim g_{UV}^2$, where $g_{UV}$ represents a coupling constant of the UV model, and we have assumed that the SMEFT operators are generated at tree-level. If we assume that the UV completion is strongly coupled we can have $g_{UV} \approx 4\pi$. In the case of the $Z$-boson form factor operators being generated by strongly coupled UV physics we would instead arrive at an effective scale of $\Lambda \approx 1$ TeV.
The corrections to $d\sigma/d m_{\ell \ell}$ induced by dimension-8 operators are given in Tables \ref{Vector8} and \ref{ZFF8}.

\section{Single coupling analysis}\label{SingleCoupling}

We now extract bounds on SMEFT coefficients from the  results of Ref. \cite{Aad:2016zzw}, which measured $p p \rightarrow \ell^+ \ell^-$, with $\ell = \{e,\mu\}$ at 8 TeV with luminosity $20.3$ fb$^{-1}$. The data are binned in twelve invariant mass bins with $m_{\ell \ell}$ varying between $m_{\ell \ell} = 116$ GeV and $m_{\ell \ell} = 1.5$ TeV. The experimental uncertainties go from 0.63\% in the smallest invariant mass bin to $17.31\%$ in the highest invariant mass bin. The uncertainty in the lower invariant mass bins is an approximately equal split between statistical and systematic errors, while in the highest bins it is dominated by statistics. An important
feature of the data set of Ref.~\cite{Aad:2016zzw} is that it was originally
intended as a SM measurement of the photon PDF, and therefore a
careful accounting of experimental errors was performed and released publicly. This is an
important point that can outweigh the improvement in constraints expected from 13 TeV collisions if those are not done with the same level of detail.

\begin{figure}
\center
 \includegraphics[width=0.7\textwidth]{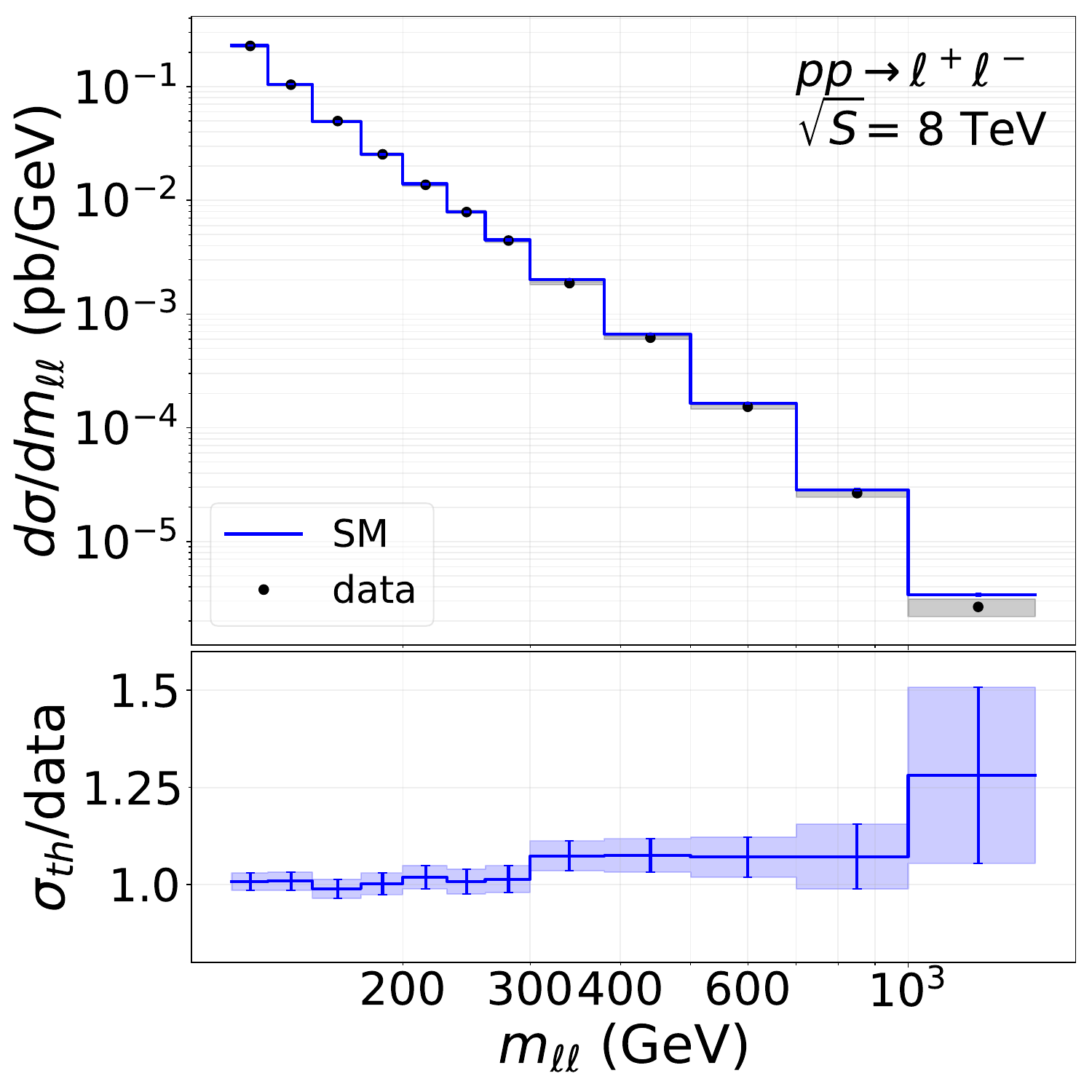}
 \caption{Comparison between the SM prediction (blue) and the measurement of the Drell-Yan invariant mass distribution of Ref. \cite{Aad:2016zzw} (black).
 In the top panel, the grey shaded area and blue error bars denote, respectively, the experimental and theoretical errors. In the bottom panel, the error on the ratio $\sigma_{\rm th}/{\rm data}$ is dominated, at high invariant mass, by the experimental uncertainties.}\label{ThData}
\end{figure}

We choose the UV scale $\Lambda = 4$ TeV, which is above the highest invariant mass bin studied in the experimental analysis, as a reference scale. We calculate the SM cross section at next-to-next-to leading order (N$^2$LO) in QCD using the $N$-jettiness subtraction method~\cite{Boughezal:2015dva,Gaunt:2015pea} as implemented in MCFM~\cite{Boughezal:2016wmq} and include next-to-leading-logarithmic (NLL) electroweak corrections~\cite{Kuhn:2005az,Hollik:2015pja}, which become important in the high invariant mass bins. The theoretical uncertainties in the SM arise from the parton distributions (PDFs), from missing higher order corrections and from uncertainties in the SM parameters.
We estimate PDF uncertainties by using the 100 members of the
\texttt{NNPDF31\_nnlo\_as\_0118} PDF set \cite{Ball:2017nwa}.
The PDF error ranges between less than 1\% and 2.8\%. PDF uncertainties between different bins are strongly correlated. We estimate the theoretical error from missing higher-order corrections by separately varying the renormalization and factorization scales in the range $m_{\ell \ell}/2 \leq \mu_{R,F} \leq 2 m_{\ell \ell}$ subject to the constraint $1/2 \leq \mu_R/\mu_F \leq 2$. To provide a conservative uncertainty estimate we vary the scales in the NLO cross section. The scale uncertainty estimated in this way ranges from 1.2\% to 3.1\% in the highest invariant mass bin. We assume that the scale uncertainty is uncorrelated between the experimental bins. The SMEFT-induced corrections are calculated at NLO in the QCD coupling constant. We have assumed no underlying hierarchy regarding the dimension-6 and dimension-8 coefficients, and rely instead upon the experimental data to determine their allowed ranges.

Figure \ref{ThData} shows the comparison between the SM prediction and the measurement of Ref. \cite{Aad:2016zzw}. We see that there is in general a very good agreement. For $m_{\ell \ell} > 300$ GeV the data lie below the SM expectation by about one sigma. 
Taking into account the experimental and theoretical correlations, for the SM cross section we find a  $\chi^2$ per degree of freedom (dof) of $11.7/12=1.05$.

\begin{figure}
\begin{centering}
 \includegraphics[width=5in]{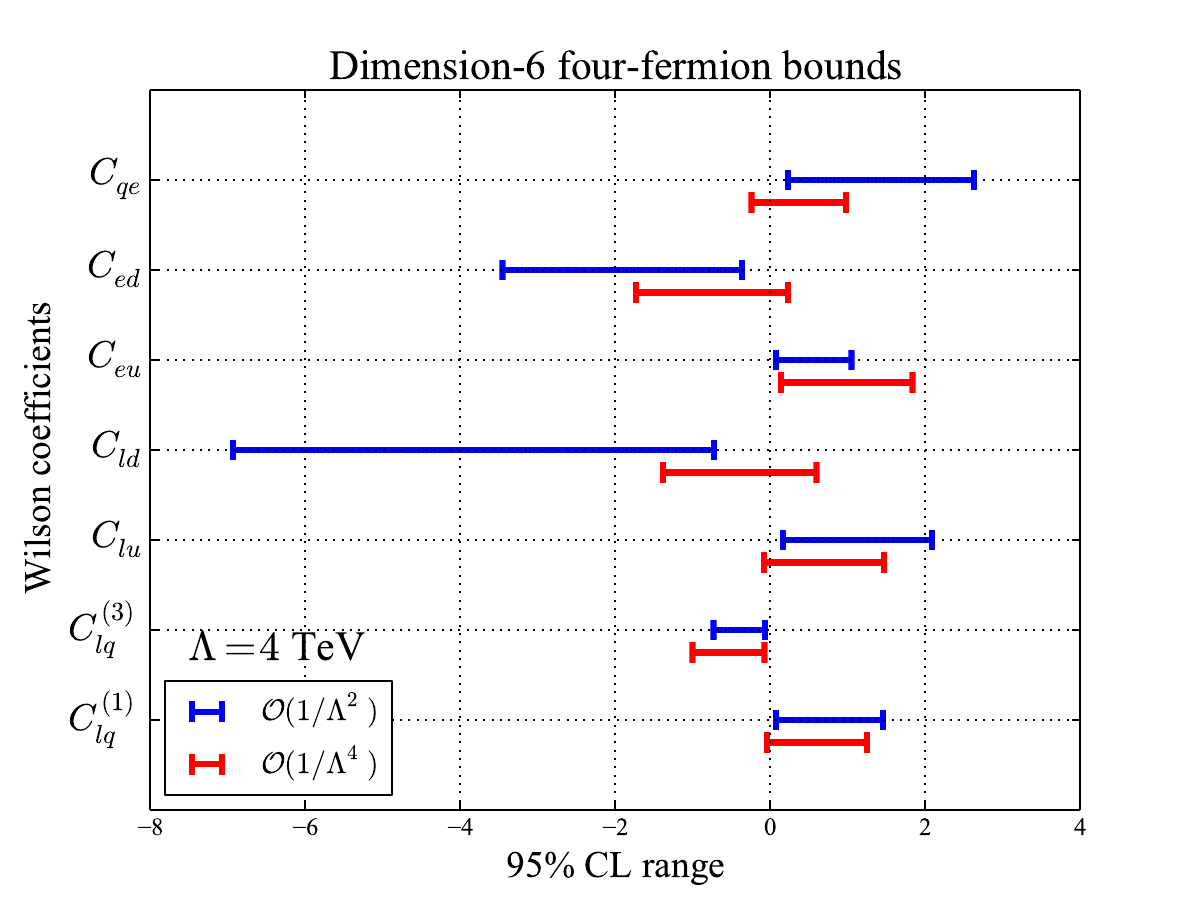}
 \caption{95\% CL intervals for the dimension-6 four-fermion operators that interfere with the SM. Both the limits obtained by considering only 
$\mathcal O(1/\Lambda^2)$ effects, as well as those including the $\mathcal O(1/\Lambda^4)$ corrections, are shown.}\label{fig:bound1}
\end{centering}
\end{figure}

\begin{figure}
\begin{centering}
 \includegraphics[width=5in]{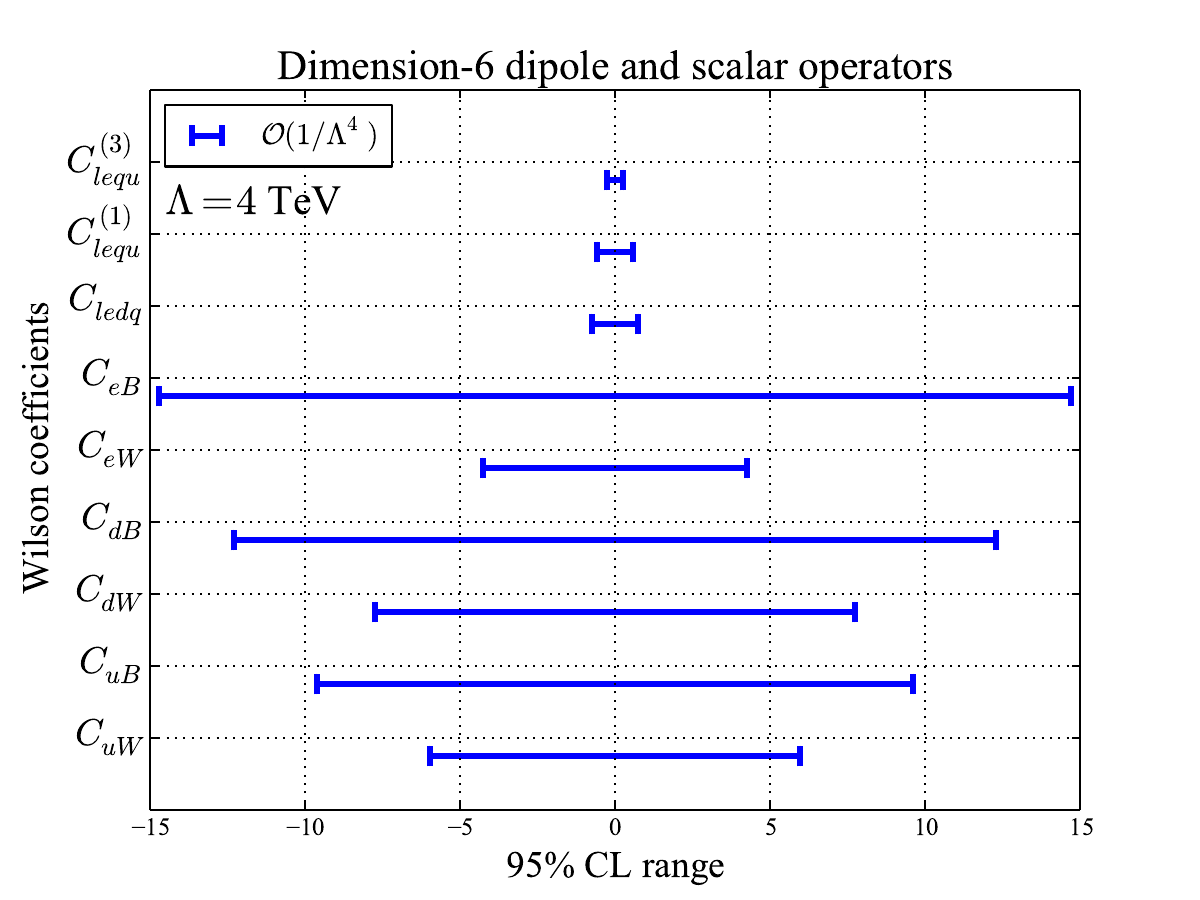}
 \caption{95\% CL intervals for dimension-6 dipole, scalar and tensor interactions that first contribute at
 $\mathcal O(1/\Lambda^4)$. The Wilson coefficients are defined at the scale $\mu_0 =1$ TeV.}\label{fig:bound2}
 \end{centering}
\end{figure}

\begin{figure}
\begin{centering}
 \includegraphics[width=5in]{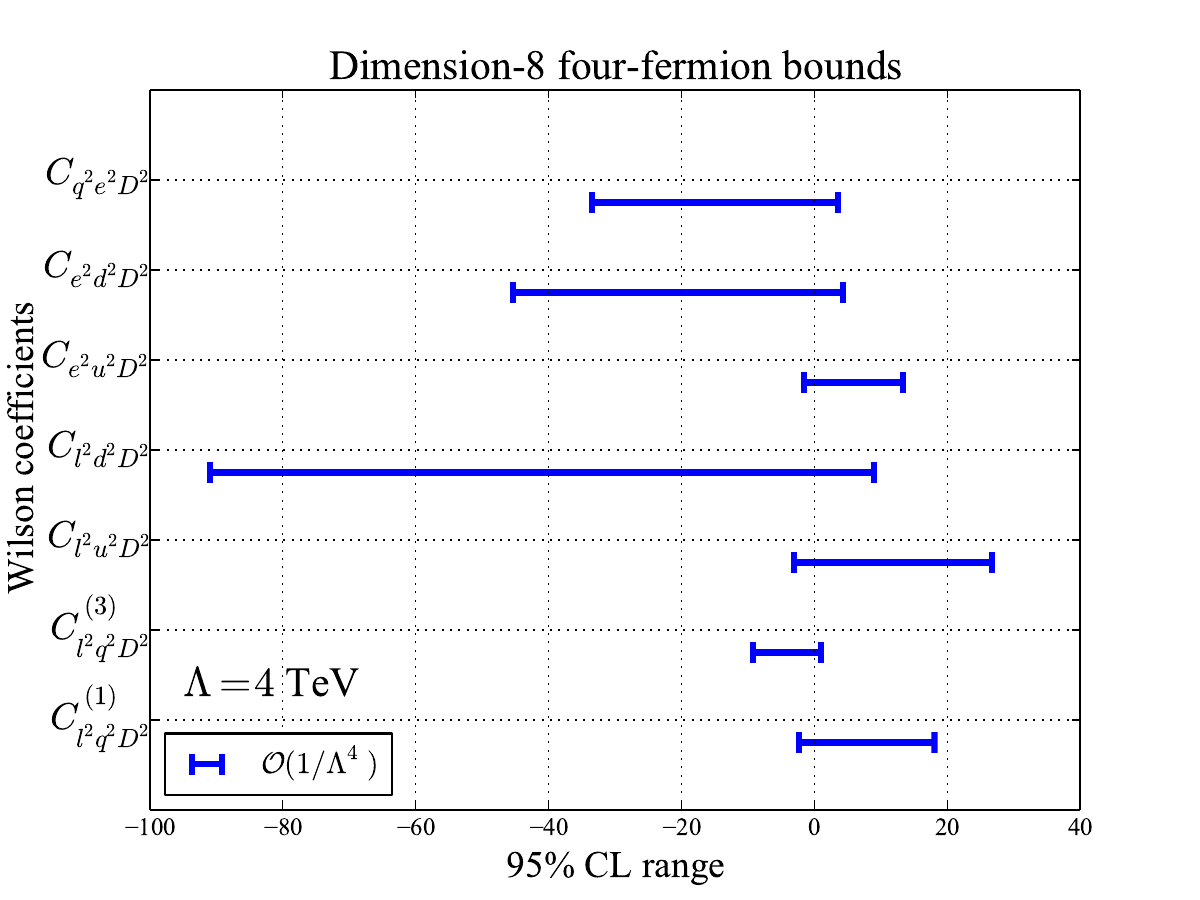}
 \caption{95\% CL intervals for the dimension-8 momentum-dependent four-fermion operators. }\label{fig:bound3}
\end{centering}
\end{figure}

\begin{figure}
\begin{centering}
 \includegraphics[width=5in]{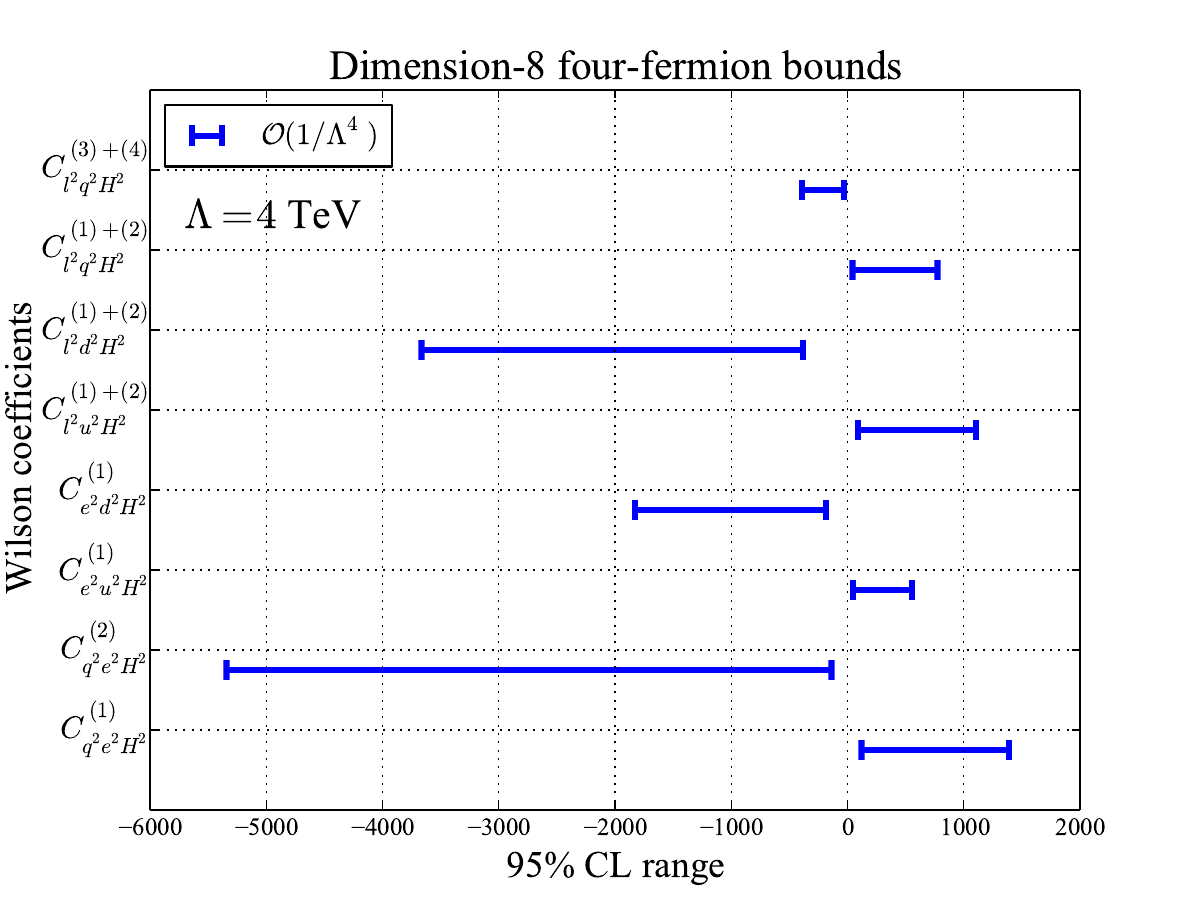}
 \caption{95\% CL intervals for the dimension-8 momentum-independent four-fermion operators. A combination such as $(1)+(2)$ in a superscript indicates that the indicated linear combination of the two relevant operators has been considered.}\label{fig:bound4}
\end{centering}
\end{figure}

\begin{figure}
\begin{centering}
 \includegraphics[width=5in]{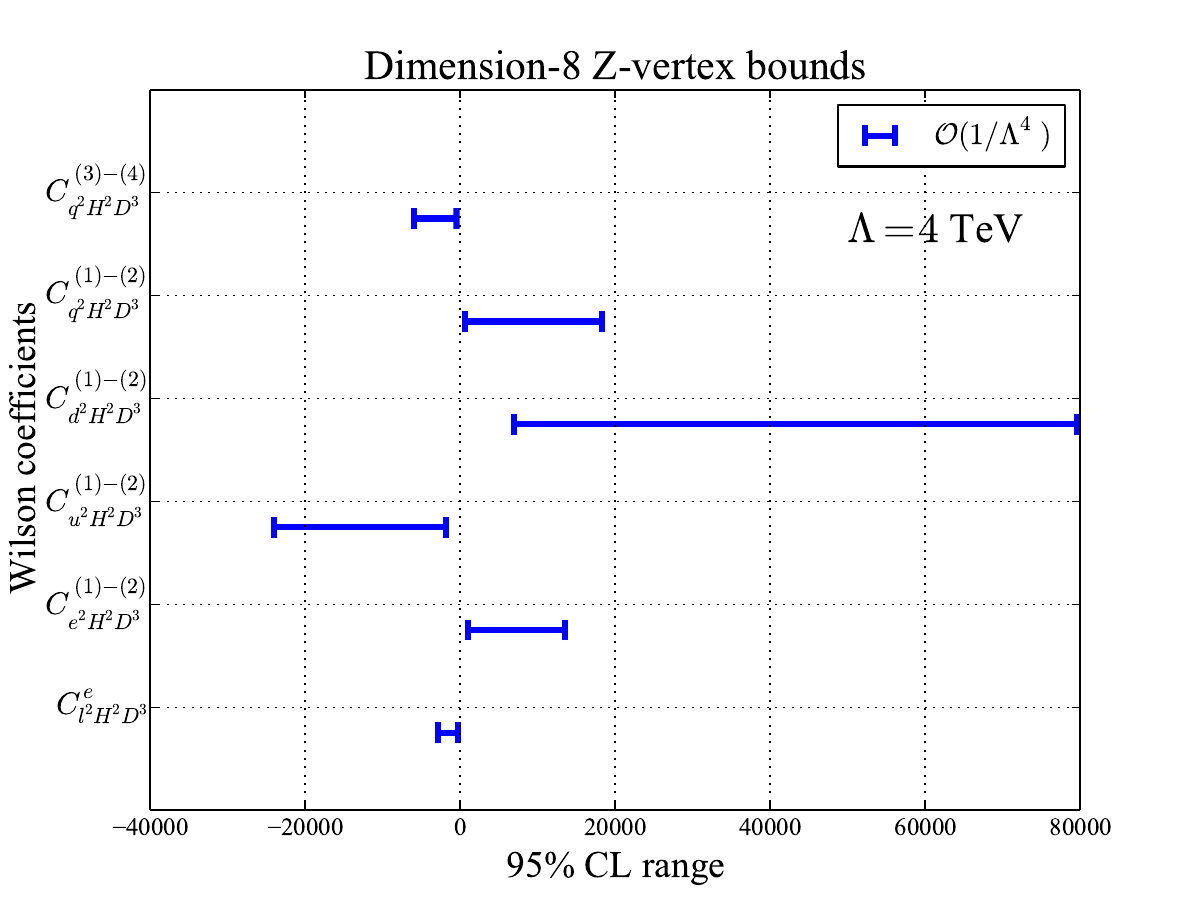}
 \caption{95\% CL intervals for a selection of dimension-8 $Z$-boson vertex corrections. A combination such as $(1)-(2)$ in a superscript indicates that the indicated linear combination of the two relevant operators has been considered.}\label{fig:bound5}
\end{centering}
\end{figure}

The bounds from turning on only a single coefficient at a time are shown in Figs.~\ref{fig:bound1}--\ref{fig:bound5}. We begin by discussing the bounds on dimension-6 four-fermion coefficients in Fig.~\ref{fig:bound1}. We compare the results obtained by keeping only $\mathcal O(1 /\Lambda^2)$ corrections with those obtained by keeping the square of dimension-6 operators that contributes $\mathcal O(1 /\Lambda^4)$ effects. In general the fits to the data are good, with a $\chi^2/$dof below one for most operators, implying that the data prefer a non-zero contribution from SMEFT operators, which interfere destructively with the SM.
The impact of $\mathcal O(1 /\Lambda^4)$ corrections are significant for most operators, with the upper and lower limits of the 95\% CL ranges shifting by factors of 2 or 3 in most cases.

We proceed to discuss next the bounds on the dipole, scalar and tensor couplings shown in Fig.~\ref{fig:bound2}. As explained previously these enter the cross section quadratically, so the bounds are symmetric around zero. The $\chi^2$ per dof is slightly over unity, indicating a reasonable fit to the data. Since these operators always increase the SM cross section and no destructive interference is possible, their inclusion does not improve the agreement with data.
The scalar and tensor operators give $\mathcal O(s^2 /\Lambda^4)$ contributions to the matrix elements squared while the dipole corrections only grow as $\mathcal O( v^2 s /\Lambda^4)$, leading to the order-of-magnitude difference in the bounds observed for these classes of operators. Defining the effective scale probed as $\Lambda/\sqrt{C_i}$ for each Wilson coefficient $C_i$, we see that scales from 4.6~TeV to 7.7~TeV are probed for scalar and tensor operators, far above the 1.5~TeV limit of the highest invariant mass bin. Scales ranging from 1.0 to 1.6~TeV are probed for the dipole operators.

It is interesting to compare these bounds with those from  flavor physics and low-energy experiments. The off-diagonal components of the scalar 
operator $C_{LedQ}$ are strongly constrained by flavor-changing-neutral-current (FCNC) decays of kaons and  $B$ mesons, such as  $B \rightarrow \ell^+ \ell^-$
or $K_{L,S} \rightarrow \ell^+ \ell^-$. These bounds can be converted into scales $\Lambda \gtrsim 150$ TeV. In the case of the scalar operators $C^{(1)}_{LeQu}$, $D_0 \rightarrow \ell^+ \ell^-$ probes the $uc$ component at the level of 20 TeV, while the limits on the tensor coupling $C^{(3)}_{LeQu}$, which contributes to $D\rightarrow \pi \ell^+ \ell^-$,  are weaker, at the TeV level.
These very stringent limits (especially on operators with $d$-type quarks) can be brought closer to the TeV scale  by assuming flavor symmetries, such as minimal flavor violation (MFV) \cite{DAmbrosio:2002vsn}. In MFV, chiral breaking operators are proportional to quark Yukawa couplings, suppressing both flavor-changing transitions but also the flavor-diagonal components that give the largest contributions to the Drell-Yan process.
Here, to avoid relying too heavily on specific flavor scenarios,
we simply account for the FCNC constraints by choosing the scalar/tensor couplings $C^{(1,3)}_{LeQu}$, $C_{LedQ}$ and the dipole couplings $C_{fB}$, $C_{fW}$, with $f\in \{u,d \}$ to be proportional to the identity in the quark mass basis\footnote{We refer to Ref. \cite{Alioli:2018ljm} for more details on the flavor conventions.}. In this case, the strongest limit on scalar operators arise from the ratios
\begin{equation}
 R_\pi = \frac{\Gamma(\pi^+ \rightarrow e^+ \nu)}{\Gamma(\pi^+ \rightarrow \mu^+ \nu)}, \quad
  R_K = \frac{\Gamma(K^+ \rightarrow e^+ \nu)}{\Gamma(K^+ \rightarrow \mu^+ \nu)},
\end{equation}
which scale as $m_e^2/m_\mu^2$ in the SM, but are not suppressed in the presence of pseudoscalar operators. Assuming flavor universality, and, in addition, that the couplings are real, one gets 
\begin{eqnarray}
-\frac{0.2}{\left(100 \, \rm TeV\right)^2 }  < \frac{1}{\Lambda^2} \left[C_{LedQ} - C^{(1)}_{LeQu} \right] < \frac{0.1}{\left(100 \, \rm TeV\right)^2}.
\end{eqnarray}
The limits on $\Lambda$ can be weakened by one order of magnitude assuming quark flavor diagonal rather than flavor universal couplings  \cite{Cirigliano:2013xha}.

For the scalar and tensor couplings, the best constraints come nuclear beta decays, $R_\pi$ and  radiative pion decays \cite{Cirigliano:2012ab,Cirigliano:2013xha,Gonzalez-Alonso:2018omy,Falkowski:2020pma}. In this case one finds \cite{Gonzalez-Alonso:2018omy,Falkowski:2020pma}
\begin{eqnarray*}
& & - \frac{0.6}{\left(4 \, \rm TeV\right)^2}  < \frac{1}{\Lambda^2}\left[C_{LedQ} + C^{(1)}_{LeQu} \right]  < \frac{0.5}{\left(4 \, \rm TeV\right)^2},
\nonumber \\
& & 
- \frac{0.5}{\left(4 \, \rm TeV\right)^2} < \frac{v^2}{\Lambda^2} \left[ C^{(3)}_{LeQu} \right]  < \frac{0.3}{\left(4 \, \rm TeV\right)^2},
\end{eqnarray*}
at 95\% CL.
These bounds are very close to those showed in Fig. \ref{fig:bound2}.
Therefore, while the linear combination constrained by $R_\pi$ and $R_K$
is out of the LHC reach,  we can conclude that for the other two linear combinations of chiral-breaking scalar and tensor coefficients there is a strong interplay between low- and high-energy searches, as already pointed out in Refs. \cite{Alioli:2018ljm,Gupta:2018qil,Gonzalez-Alonso:2018omy,Falkowski:2020pma}.
Similar conclusions apply to the real part of flavor-diagonal dipole operators. 
Electric dipole moments put strong constraints on the imaginary part of the coefficients of flavor-diagonal chiral-breaking operators, so that for these the LHC is never competitive.

Finally we proceed to discuss the dimension-8 operator bounds in Figs.~\ref{fig:bound3},~\ref{fig:bound4}, and~\ref{fig:bound5}. The first plot discusses operators of the form $\mathcal L_{\psi^4 D^2}$ that give momentum-dependent four-fermion corrections that scale as $\mathcal O(s^2/\Lambda^4)$, the second gives momentum-independent four-fermion corrections from $\mathcal L_{\psi^4 H^2}$ that scale as $\mathcal O( v^2 s/\Lambda^4)$, and the third gives momentum-dependent $V\bar{f}f$ vertex corrections, which also scale as $\mathcal O(v^2 s/\Lambda^4)$. These different scalings lead to vast differences in the effective scales probed for each type of operator. Defining the effective scale as $\Lambda/\sqrt[\leftroot{-2}\uproot{2}4]{C_i}$ for these dimension-8 terms, UV scales ranging from 1.3 to 4.3~TeV are reached for operators of the $\mathcal L_{\psi^4 D^2}$ class. Given that these are pure dimension-8 effects that appear first at $\mathcal O(1/\Lambda^4)$ this is striking. Since these scales approach those of the dimension-6 effects, these operators cannot be safely neglected in fits to the data. We study this point further in the next section. Bounds on operators from the $\mathcal L_{\psi^4 H^2}$ class are weaker, with the maximum scale probed reaching 1.7~TeV at most. The scales associated with the vertex corrections do not reach 1 TeV, and these corrections can be safely neglected in fits compared to the larger corrections arising from dimension-6 effects. 

We used in this analysis 8 TeV data, for which detailed information
about the experimental errors and their correlations is available. 
Including the full correlation matrix is particularly important for the constraints on operators that interfere with the SM. Since for $m_{\ell \ell} > 300$ GeV all bins lie below the SM expectation,
neglecting the experimental correlations leads to 
a better SM fit and 
changes the smaller side of the bounds 
in Figs. \ref{fig:bound1}, \ref{fig:bound3}, \ref{fig:bound4} and \ref{fig:bound5} by up to a factor of two.
The bounds on scalar, tensor and dipole operators, which do not interfere with the SM, are only marginally affected by neglecting correlations.
ATLAS and CMS have also published searches for resonant and non-resonant phenomena in high-mass dilepton and lepton plus missing energy final states
based on 40 fb$^{-1}$ and 139 fb$^{-1}$ of 13 TeV data \cite{ATLAS:2017fih,CMS:2018ipm,CMS:2018hff,ATLAS:2019lsy,ATLAS:2019erb,ATLAS:2020yat,CMS:2021ctt}.
While these datasets play an important role in further constraining the SMEFT expansion, they are published with less detailed error information, and the extraction of reliable bounds requires a detailed detector simulation not available to theorists. We note that 
the search for contact interactions in Ref. \cite{ATLAS:2020yat} considers a subset of the dimension-6 operators that we included, and uses a signal region of  $2 < m_{\ell \ell} < 6$ TeV. 
The uncertainties on the background in this region are quite significant, and the limits on the new physics scale $\Lambda$, once converted into the conventions of Eq. \eqref{eq:fourfermion},
are about $\Lambda \sim 7$ TeV, stronger by only a factor of approximately 1.5 compared to our analysis. This again highlights the importance that precise data at all energies can have on SMEFT analyses.

\section{Multiple couplings scenarios}\label{Multiple}

We study in this section the impact of turning on several dimension-6 and dimension-8 SMEFT operators at the same time.
Since angular information can in principle disentangle operators with different helicities \cite{Alioli:2018ljm}, we consider one specific helicity channel, with right-handed $u$ quarks and right-handed electrons.
The dimension-6 operator that contributes	 in this channel is
$C_{eu}$. At dimension eight, 
we turn on $C_{e^2 u^2 H^2}$,
the derivative operator $C_{e^2 u^2 D^2}$, and one momentum-dependent correction to the $Z$-vertex, 
$C^{(1)}_{u^2 H^2 D^3} - C^{(2)}_{u^2 H^2 D^3}$ (we note that $C^{(1)}_{u^2 H^2 D^3} - C^{(2)}_{u^2 H^2 D^3}$ also contributes to the helicity channel with right-handed quarks and left-handed electrons). With these four couplings, we find a best fit $\chi^2/{\rm dof} = 5.8/8$, indicating that multiple couplings do not significantly improve the fits. 
The 95\% CL limits on the four SMEFT operators we considered are shown in Table \ref{marginalized}.
In the two leftmost columns we give the limits obtained in the single coupling hypothesis, truncating the EFT expansion at dimension-6 and dimension-8, respectively. In the third column, we show the 95\% CL limits obtained by marginalizing over the remaining three couplings.  
In this case, the  correlation matrix of the four couplings $(C_{eu},C_{e^2 u^2 D^2},C_{e^2 u^2 H^2}, C^{(1)}_{u^2 H^2 D^3} - C^{(2)}_{u^2 H^2 D^3})$ is given by
\begin{equation}
 {\rm corr} = \left(\begin{array}{cccc}
                     1.00   & 0.995  &    -0.91 & -0.02 \\
                     0.995  & 1.00   &    -0.92 & -0.02  \\
                     -0.91  & -0.92  &  1.00 & 0.42  \\
                     -0.02  & -0.02  &  0.42 & 1.00  \\
                \end{array} \right),
\end{equation}
indicating strong correlations between dimension-6 and dimension-8 four-fermion operators. 
The rightmost column show marginalized bounds, but with the coefficients of dimension-8 operators allowed to vary between $\pm 256$, so that the effective scale of the operators does not go below 1 TeV. 

\begin{table}
\center 
 \begin{tabular}{|c || c ||c  || c  | c ||}
\hline 
& {dim-6} & {single coupling} & {marginalized} & marginalized$^*$  \\
\hline
$C_{eu}$    &  $\left[0.08,1.0\right]$     & $\left[0.1,1.8\right]$  &  $\left[-39,39\right]$ & $[-0.6,2.4]$\\
$C_{e^2u^2 D^2}$ & -- & $\left[-1.5,13\right]$  &  $\left[-17,9.2 
\cdot 10^3 \right]$  & $[-14,18]$\\
$C_{e^2u^2 H^2}$ & --& $\left[45,555\right]$  &  $\left[-1.9,1.2\right] \cdot 10^{4}$  & $[-256,256]$\\
$C^{(1)-(2)}_{u^2 H^2 D^3}$ & -- & $[-24 ,-1.8 ] \cdot 10^{3}$
& $[-1.2 , 1.8 ] \cdot 10^{5}$ & $[-256,256]$
\\
\hline 
\end{tabular}
\caption{95\% CL intervals for four operators with right-handed $u$ quarks. The first and second
columns show the bounds obtained assuming that only one operator is on at a time,
and truncating the EFT expansion at dimension-6 and dimension-8, respectively. The third column shows the limits on a given coefficient obtained marginalizing over the other three.
In the rightmost column, we marginalize over the couplings, but allow dimension-8 coefficients to vary between $\pm 256$, corresponding to an effective scale of 1 TeV.
}\label{marginalized}
\end{table}
	
From Table \ref{marginalized}, we see that the bounds on the dimension-6 operator $C_{eu}$ can be weakened by turning on dimension-8 operators with arbitrary coefficients. In the case that all couplings are allowed to vary freely without enforcing consistency of the EFT expansion for all non-zero couplings, the bounds on $C_{eu}$ are weakened by more than an order of magnitude. Even if we constrain the dimension-8 operators to have coefficients compatible with the EFT expansion, the derivative operator
$C_{e^2u^2 D^2}$ plays an important role, 
and weakens the bounds on $C_{eu}$ by a factor of 2 compared to the single coupling analysis truncated at $\mathcal O(1/\Lambda^2)$. We conclude that fits to the Drell-Yan data that truncate to dimension-6 operators only can be misleading by a significant amount.

\section{Conclusions}\label{Conclusion}

In this manuscript we have studied the Drell-Yan process at $\mathcal O(1/\Lambda^4)$ in the SMEFT, including effects from both the square of dimension-6 operators and genuine dimension-8 effects. Our calculation of the SMEFT contributions to this process includes NLO QCD corrections through $\mathcal O(\alpha_s / \Lambda^4)$. It is missing only effects from dimension-8 operators containing explicit gluon fields that are not enhanced in the soft or collinear limits. We have found that corrections from $\mathcal O(1/\Lambda^4)$ are significant, with the terms quadratic in the dimension-6 Wilson coefficients becoming as important as the linear effects far below the UV scale $\Lambda$. Energy-dependent dimension-8 four-fermion operators whose effects scale as $s^2/\Lambda^4$ in the high-energy limit also become nearly as large as the dimension-6 terms for $s \ll \Lambda$.

To illustrate the impact of these findings we perform fits to the ATLAS high-mass data from Ref.~\cite{Aad:2016zzw}. Our fits include the full experimental correlated errors, as well as the SM Drell-Yan cross section calculation through NNLO in QCD and NLL in the electroweak coupling constant. Inclusion of the quadratic dimension-6 effects can shift the limits on the relevant Wilson coefficients by factors of 2-3. The dimension-8 effects further shift the bounds by additional large factors. Our findings clearly show that truncation of the SMEFT expansion to $\mathcal O(1/\Lambda^2)$ does not properly account for the SMEFT effects on the Drell-Yan process. We note that improved experimental precision in the higher invariant mass bins would reduce the impact of $\mathcal O(1/\Lambda^4)$ effects.  In the last bin the ATLAS data considered here allows for a 25\% deviation from the SM. Improving this precision would reduce the allowed parameters to a region where the SMEFT expansion converges more quickly.

Although the inclusion of the full dimension-8 corrections in the SMEFT may at first appear to be a daunting task, our results show that only the subset of dimension-8 operators consisting of two-derivative four-fermion interaction must be included. Other categories of corrections, including energy-independent dimension-8 four-fermion operators whose effects scale as $v^2 s/\Lambda^4$ and $Z$-boson vertex corrections, can be neglected given the current data precision. We believe that our findings provide a solid foundation for future analyses of LHC Drell-Yan measurements within the SMEFT framework. 
To this goal, on the experimental side it will be important to have access to more differential distributions, including rapidity and angular distributions, at high invariant mass. These measurements will allow to disentangle the flavor and helicity structure of dimension-6 and dimension-8 operators, reducing the degeneracies that affect the dilepton invariant mass distribution.
On the theoretical side, a consistent dimension-8 fit will need to include the 
``positivity'' constraints that can be inferred by fundamental principles of quantum field theory
\cite{Adams:2006sv,Zhang:2018shp,Zhang:2020jyn,Bi:2019phv,Remmen:2019cyz,Remmen:2020vts,Yamashita:2020gtt,Trott:2020ebl,Fuks:2020ujk}.
While the most naive elastic positivity constraints do not apply to the operators in the class $\mathcal L_{\psi^4 D^2}$  that are most relevant to Drell-Yan 
\cite{Remmen:2020vts}, a more detailed analysis of elastic positivity
and extremal positivity bounds is necessary \cite{Remmen:2020vts,Yamashita:2020gtt}.

\acknowledgments
We thank S. Alioli for many stimulating conversations and for help with the implementation of dimension-8 operators in \texttt{POWHEG}. 
R. B. is supported by the DOE contract DE-AC02-06CH11357. E.~M. is supported  by the US Department of Energy through  
the Office of Nuclear Physics  and  the  
LDRD program at Los Alamos National Laboratory. Los Alamos National Laboratory is operated by Triad National Security, LLC, for the National Nuclear Security Administration of U.S.\ Department of Energy (Contract No. 89233218CNA000001). F. P. is supported by the DOE grants DE-FG02-91ER40684 and DE-AC02-06CH11357. This research was supported in part through the computational resources and staff contributions provided for the Quest high performance computing facility at Northwestern University which is jointly supported by the Office of the Provost, the Office for Research, and Northwestern University Information Technology. We thank S.~Hamoudou for pointing out typos in the Appendix of the original version of this manuscript.

\appendix

\section{Vertex parameterizations}
\label{sec:Vparam}

In order to discuss exactly which combinations of Wilson coefficients can be probed in the Drell-Yan process, it is helpful to introduce general parameterizations for the gauge boson-fermion vertices and for the four-fermion interactions. We begin with the four-fermion interactions. We parameterize the four-fermion interaction as (quark momenta
incoming, lepton momenta outgoing):
\begin{equation}
\begin{split}  
  -iV_{q_3\bar{q}_4l_1\bar{l}_2} &= V^{(6)}_{LL}(\gamma^{\mu} P_L)^q
  (\gamma_{\mu}P_L)^l+V^{(6)}_{RR}(\gamma^{\mu} P_R)^q (\gamma_{\mu}P_R)^l
+V_{RL}^{(6)}(\gamma^{\mu} P_R)^q (\gamma_{\mu}P_L)^l\nonumber \\
&+V^{(6)}_{LR}(\gamma^{\mu} P_L)^q (\gamma_{\mu}P_R)^l
  +S_a^{(6)} (P_L^q P_L^l + P_R^q P_R^l )  +S_b^{(6)} (P_R^q P_L^l + P_L^q P_R^l ) \nonumber \\ &+ T^{(6)} \left[ (\sigma^{\mu\nu}
  P_L)^q  (\sigma_{\mu\nu} P_L)^l+ (\sigma^{\mu\nu} P_R)^q
  (\sigma_{\mu\nu} P_R)^l\right] \nonumber \\
&+(p_1+p_2)\cdot (p_3+p_4) \left[ V^{(8)}_{LL}(\gamma^{\mu} P_L)^q
  (\gamma_{\mu}P_L)^l+V^{(8)}_{RR}(\gamma^{\mu} P_R)^q (\gamma_{\mu}P_R)^l
  \right.
\nonumber \\ & \left.+V^{(8)}_{RL}(\gamma^{\mu} P_R)^q (\gamma_{\mu}P_L)^l+V^{(8)}_{LR}(\gamma^{\mu} P_L)^q
  (\gamma_{\mu}P_R)^l\right]
  \end{split}
\end{equation}  
As explained previously we have neglected operators of the type $\bar q_L  \gamma^{(\mu} \overleftrightarrow D^{\nu )} q_L \, \bar{\ell}_L \gamma_{(\mu} \overleftrightarrow{D}_{\nu)} \ell_L$.  We note that this parameterization holds for both up and down type quarks.
The coefficients $S^{(6)}$, $T^{(6)}$, and the $V^{(8)}$ all
begin contributing to the cross section at $\mathcal O(1/\Lambda^4)$.  The $V^{(6)}$ begin at $\mathcal O(1/\Lambda^2)$. They have the following expansion:
\begin{equation}
V^{(6)} = \frac{1}{\Lambda^2} V^{(6a)}+\frac{1}{\Lambda^4} V^{(6b)}.
\label{eq:Vdef}
\end{equation}  

It is straightforward to use the operators listed in the previous section to determine these vertex factors in terms of Wilson coefiicients. We begin with the scalar and tensor couplings defined in $\mathcal L_{\psi^4}$:
\begin{eqnarray}
S^{(6)}_{a,u} &=& -C_{\ell equ}^{(1)}/\Lambda^2, \nonumber \\
S^{(6)}_{b,u} &=& 0, \nonumber \\
T^{(6)}_u &=& -C_{\ell equ}^{(3)}/\Lambda^2, \nonumber \\
S^{(6)}_{a,d} &=& 0, \nonumber \\
S^{(6)}_{b,d} &=& C_{\ell edq}/\Lambda^2, \nonumber \\
T^{(6)}_d &=& 0
\end{eqnarray}  
All six scalar and tensor Wilson coefficients can in principle be determined since they appear in different vertex factors, and lead to different angular dependences in the cross section. The matrix element squared for these interactions, which only appears interfered with itself, scales as $\mathcal O(s^2/\Lambda^4)$.

We next consider the $V^{(8)}$ vertex factors which come from $\mathcal L_{\psi^4D^2}$.  These can be written in terms of Wilson coefficients as
\begin{eqnarray} 
  V^{(8)}_{LL,u} &=&  (C^{(1)}_{\ell^2q^2D^2}- C^{(3)}_{\ell^2q^2D^2})/\Lambda^4,\nonumber \\
  V^{(8)}_{LL,d} &=&  (C^{(1)}_{\ell^2q^2D^2}+ C^{(3)}_{\ell^2q^2D^2})/\Lambda^4,\nonumber \\
  V^{(8)}_{RR,u} &=&   C^{(1)}_{e^2u^2D^2}/\Lambda^4, \nonumber \\
  V^{(8)}_{RR,d} &=&   C^{(1)}_{e^2d^2D^2}/\Lambda^4, \nonumber \\
  V^{(8)}_{RL,u} &=&   C^{(1)}_{\ell^2u^2D^2}/\Lambda^4, \nonumber \\
  V^{(8)}_{RL,d} &=&   C^{(1)}_{\ell^2d^2D^2}/\Lambda^4, \nonumber \\
  V^{(8)}_{LR,u} &=&   C^{(1)}_{q^2e^2D^2}/\Lambda^4, \nonumber \\
  V^{(8)}_{LR,d} &=&  C^{(1)}_{q^2e^2D^2}/\Lambda^4.
  \label{eq:V8def}
\end{eqnarray}  
All seven Wilson coefficients appear in separate vertex factors. We note that these contributions lead to different energy dependences than the $V^{6}$ vertex correction factors defined in Eq.~({\ref{eq:Vdef}}), and the effects of these two classes of operators can in principle be disentangled in the Drell-Yan process. Contributions from these vertex factors scale as $\mathcal O(s^2/\Lambda^4)$ in the high-energy limit.

We now proceed to the momentum-independent four-fermion operators that first contribute at $\mathcal O(1/\Lambda^2)$.  The first terms in their expansion take the forms
\begin{eqnarray}\label{Vff1}
      V^{(6a)}_{LL,u} &=& C_{\ell q}^{(1)}-C_{\ell q}^{(3)},\nonumber \\
      V^{(6a)}_{RR,u} &=& C_{eu},\nonumber \\
      V^{(6a)}_{LR,u} &=& C_{qe},\nonumber \\
      V^{(6a)}_{RL,u} &=& C_{\ell u},\nonumber \\
      V^{(6a)}_{LL,d} &=& C_{\ell q}^{(1)}+C_{\ell q}^{(3)},\nonumber \\
      V^{(6a)}_{RR,d}& =& C_{ed},\nonumber \\
      V^{(6a)}_{LR,d} &=& C_{qe},\nonumber \\
      V^{(6a)}_{RL,d} &=& C_{\ell d}.
\end{eqnarray}
The $\mathcal O(1/\Lambda^4)$ terms in the expansion come from $\mathcal L_{\psi^4 H^2}$, and take the form
\begin{eqnarray}\label{Vff2}
    V^{(6b)}_{RR,u} &= &\frac{v^2}{2}C_{e^2u^2H^2} ,\nonumber \\
    V^{(6b)}_{RR,d} &=& \frac{v^2}{2}C_{e^2d^2H^2} ,\nonumber \\
    V^{(6b)}_{RL,u} &=& \frac{v^2}{2}(C^{(1)}_{\ell^2u^2H^2}+C^{(2)}_{\ell^2u^2H^2}) ,\nonumber \\
    V^{(6b)}_{RL,d} &=& \frac{v^2}{2}(C^{(1)}_{\ell^2d^2H^2}+C^{(2)}_{\ell^2d^2H^2}) ,\nonumber \\
    V^{(6b)}_{LR,u} &=& \frac{v^2}{2}(C^{(1)}_{q^2e^2H^2}-C^{(2)}_{q^2e^2H^2}) ,\nonumber \\
    V^{(6b)}_{LR,d} &=& \frac{v^2}{2}(C^{(1)}_{q^2e^2H^2}+C^{(2)}_{q^2e^2H^2}) ,\nonumber \\
    V^{(6b)}_{LL,u} &=& \frac{v^2}{2}(C^{(1)}_{\ell^2q^2H^2}+C^{(2)}_{\ell^2q^2H^2}-C^{(3)}_{\ell^2q^2H^2}-C^{(4)}_{\ell^2q^2H^2}) ,\nonumber \\
    V^{(6b)}_{LL,d} &=& \frac{v^2}{2}(C^{(1)}_{\ell^2q^2H^2}+C^{(2)}_{\ell^2q^2H^2}+C^{(3)}_{\ell^2q^2H^2}+C^{(4)}_{\ell^2q^2H^2}).
\end{eqnarray}
Only eight linear combinations of the twelve coefficients in $\mathcal L_{\psi^4 H^2}$ appear in the vertex factors. These corrections scale as $\mathcal O(v^2s/\Lambda^4)$. They do not grow with energy as quickly as the $V^{(8)}$ contributions. We note that the overall $\Lambda$ dependence of these corrections has been extracted following the definition in Eq.~(\ref{eq:Vdef}).
The remaining four linear combination appear in vertices with two neutrinos and right- or left-handed $u$ and $d$ quarks, and are therefore difficult to probe.

We now proceed to parameterize the corrections to the photon and $Z$-boson vertices. We express these vertices as
\begin{eqnarray}
iV_{A\bar{f}_if_i}&=&  \left\{\bar{e} Q_i\gamma^{\mu} -i D_{A_i}
                 \sigma^{\mu\nu} p_{A\nu} \right\} \nonumber \\
iV_{Z\bar{f}_if_i}  &=&  \left\{V_{ZL_i}P_L+V_{ZR_i}P_R + p_Z^2
                   \left[W_{ZL_i}P_L+W_{ZR_i}P_R\right]
                   -iD_{Z_i}
                 \sigma^{\mu\nu} p_{Z\nu} \right\}.
\end{eqnarray}  
$P_{L,R}$ are the standard left and right-handed projection operators, and $Q_i$ is the electric charge of fermion $i$. The normal SM couplings receive corrections in SMEFT, and must be expanded in $\kappa=1/\Lambda^2$:
\begin{eqnarray}
  \bar{e} &=& \hat{e}+\kappa \bar{e}_1+\kappa^2\bar{e}_2, \nonumber \\
  \bar{g}_Z &=& \hat{g}+\kappa \bar{g}_{Z1}+\kappa^2\bar{g}_{Z2}, \nonumber \\
  \bar{s}^2_W &=& \hat{s}^2_W+\kappa \bar{s}^2_{W1}+\kappa^2\bar{s}^2_{W2}, \nonumber \\
V_{ZL_i} &=& V_{ZL0_i}+\kappa  V_{ZL1_i}+\kappa^2 
                V_{ZL2_i}, \nonumber \\
V_{ZR_i} &=& V_{ZR0_i}+\kappa  V_{ZR1_i}+\kappa^2  
                V_{ZR2_i}  .
\end{eqnarray}  
We have shown as well the expansion of the momentum-independent $Z$-boson vertex factors. We note that the dipole corrections, as well as the momentum-dependent vertex corrections factors $W_{Z}$, contribute first at $\mathcal O(\kappa^2)$. The expansion of the input parameters in $\kappa$ has been studied to all orders in the SMEFT~\cite{Helset:2020yio,Hays:2020scx}.

We can write all of these variables in terms of the input parameters
and Wilson coefficients.  We begin with the dipole terms:
\begin{eqnarray}
D_{Z_l} = \sqrt{2} \frac{\hat{v}}{\Lambda^2} \left(
  \hat{s}_W C_{eW}-\hat{c}_WC_{eB}\right),\nonumber \\
D_{Z_l} = \sqrt{2} \frac{\hat{v}}{\Lambda^2} \left(
  \hat{c}_WC_{eW}+\hat{s}_WC_{eB}\right),\nonumber \\  
D_{A_u} = \sqrt{2} \frac{\hat{v}}{\Lambda^2} \left(
  -\hat{s}_WC_{uW}-\hat{c}_WC_{uB}\right),\nonumber \\
D_{Z_u} = \sqrt{2} \frac{\hat{v}}{\Lambda^2} \left(
  -\hat{c}_WC_{uW}+\hat{s}_WC_{uB}\right),\nonumber \\  
D_{A_d} = \sqrt{2} \frac{\hat{v}}{\Lambda^2} \left(
  \hat{s}_WC_{dW}-\hat{c}_WC_{dB}\right),\nonumber \\
D_{Z_d} = \sqrt{2} \frac{\hat{v}}{\Lambda^2} \left(
  \hat{c}_WC_{dW}+\hat{s}_WC_{dB}\right).
\end{eqnarray}  
The ${\cal O}(\kappa^0)$ weak mixing angle and Higgs vev are defined in the $(G_F,M_W,M_Z)$ scheme used here as
\begin{eqnarray}
  \hat{s}_W &=& 1-\frac{M_W^2}{M_Z^2}, \nonumber \\
  \hat{v}^2 &=& \frac{1}{\sqrt{2}G_F}.
\end{eqnarray}  
Since the dipole terms first contribute at
${\cal O}(\kappa^2)$ we have replaced the couplings with
their ${\cal O}(\kappa^0)$ values.  The following six Wilson
coefficients contribute to the dipole terms:
\begin{equation}
C_{eW},C_{eB}, C_{uW},C_{uB}, C_{dW},C_{dB}.
\end{equation}  
These corrections scale as $\mathcal O(v^2s/\Lambda^4)$ in the high-energy limit.

We now present the momentum-dependent vertex corrections.  These can be written as
\begin{eqnarray}
W_{ZL_l} &=& \frac{\hat{g}_Z\hat{v}^2}{4 \Lambda^4}  C^{(e)}_{\ell^2 H^2 D^3}, \nonumber \\
W_{ZR_l} &=& \frac{\hat{g}_Z\hat{v}^2}{4 \Lambda^4} \left[  C^{(1)}_{e^2 H^2 D^3}
-  C^{(2)}_{e^2 H^2 D^3}\right], \nonumber \\
W_{ZL_u} &=& \frac{\hat{g}_Z\hat{v}^2}{4 \Lambda^4} \left[  C^{(1)}_{q^2 H^2 D^3}
-  C^{(2)}_{q^2 H^2 D^3} -
   C^{(3)}_{q^2 H^2 D^3}
+  C^{(4)}_{q^2 H^2 D^3} 
\right] \nonumber \\
W_{ZR_u} &=& \frac{\hat{g}_Z\hat{v}^2}{4 \Lambda^4} \left[  C^{(1)}_{u^2 H^2 D^3}
-  C^{(2)}_{u^2 H^2 D^3}\right] \nonumber \\
W_{ZL_d} &=& \frac{\hat{g}_Z\hat{v}^2}{4 \Lambda^4} \left[  C^{(1)}_{q^2 H^2 D^3}
-  C^{(2)}_{q^2 H^2 D^3} +
   C^{(3)}_{q^2 H^2 D^3}-
  C^{(4)}_{q^2 H^2 D^3} 
\right], \nonumber \\
W_{ZR_d} &=& \frac{\hat{g}_Z\hat{v}^2}{4 \Lambda^4} \left[ C^{(1)}_{d^2 H^2 D^3}
-  C^{(2)}_{d^2 H^2 D^3}\right].
\end{eqnarray}  
%
where $C^{(e)}_{\ell H^2 D^3}$ was defined in Eq. \eqref{Cel2H2D3}. 
The ${\cal O}(\kappa^0)$ $Z$-coupling is
\begin{equation}
\hat{g}_Z = 2\times 2^{1/4}M_Z\sqrt{G_F},
\end{equation}  
Since these corrections contribute first at $\mathcal O(\kappa^2)$ we can use the leading-order expressions for the $Z$-boson coupling. We see that these couplings depend on the following six combinations of
Wilson coefficients:
%
%
\begin{equation}
   \begin{split}
& C^{(1)}_{\ell^2 H^2 D^3} -  C^{(2)}_{\ell^2 H^2 D^3} + 
 C^{(3)}_{\ell^2 H^2 D^3}  - C^{(4)}_{\ell^2 H^2 D^3},
 C^{(1)}_{e^2 H^2 D^3} -  C^{(2)}_{e^2 H^2 D^3} \nonumber \\
 & C^{(1)}_{q^2 H^2 D^3} -  C^{(2)}_{q^2 H^2 D^3},  
 C^{(3)}_{q^2 H^2 D^3}  - C^{(4)}_{q^2 H^2 D^3}, C^{(1)}_{u^2 H^2 D^3} -  C^{(2)}_{u^2 H^2 D^3}, C^{(1)}_{d^2 H^2 D^3} -  C^{(2)}_{d^2 H^2 D^3}.
\end{split}
\end{equation}
There are 14 Wilson coefficients in total, so multiple flat directions
appear in the parameter space.  We enumerate the eight directions that cannot be probed in neutral-current Drell-Yan below: 
\begin{equation}
   \begin{split}
& C^{(1)}_{\ell^2 H^2 D^3} -  C^{(2)}_{\ell^2 H^2 D^3} - 
 C^{(3)}_{\ell^2 H^2 D^3}  + C^{(4)}_{\ell^2 H^2 D^3}, \nonumber \\
&  C^{(1)}_{\ell^2 H^2 D^3} +  C^{(2)}_{\ell^2 H^2 D^3},\, 
 C^{(3)}_{\ell^2 H^2 D^3} +  C^{(4)}_{\ell^2 H^2 D^3},\, 
 C^{(1)}_{e^2 H^2 D^3} +  C^{(2)}_{e^2 H^2 D^3} \nonumber \\
 & C^{(1)}_{q^2 H^2 D^3} +  C^{(2)}_{q^2 H^2 D^3},  
 C^{(3)}_{q^2 H^2 D^3}  + C^{(4)}_{q^2 H^2 D^3}, C^{(1)}_{u^2 H^2 D^3} +  C^{(2)}_{u^2 H^2 D^3}, C^{(1)}_{d^2 H^2 D^3} +  C^{(2)}_{d^2 H^2 D^3}.
\end{split}
\end{equation} 
%
The first linear combination is in principle accessible in $ p p \rightarrow \nu \nu$ or $Z\rightarrow \nu \nu$. Probing the remaining combinations requires processes with multiple Higgs and gauge bosons. The $W_Z$ terms induce corrections to the Drell-Yan cross section 
that scale as $\mathcal O (v^2s/\Lambda^2)$ in the high-energy limit.

We now proceed to study the momentum-independent $Z$-boson vertex
factors.  The ${\cal O}(\kappa^0)$ pieces are given by
\begin{eqnarray}
V_{ZL0_i} &=& \hat{g}_Z \left( I^3_i-Q_i \hat{s}_W^2 \right),
                 \nonumber \\
V_{ZR0_i} &=& \hat{g}_Z \left(-Q_i \hat{s}_W^2 \right).
\end{eqnarray}  
The ${\cal O}(\kappa^1)$ pieces contain two distinct contributions:
the expansion of the couplings $\hat{g}_Z$ and $\hat{s}_W^2$, and the
explicit dimension-6 vertices.  The explicit vertices can be obtained from Ref.~\cite{Murphy:2020rsh}:
\begin{eqnarray}
V_{ZL1_l} &=& \bar{g}_{Z1} \left( I^3_e-Q_e \hat{s}_W^2 \right)
                 -\hat{g}_Z Q_e \bar{s}_{W1}^2-\frac{\hat{g}_Z
                 \hat{v}^2}{2}(C^{(1)}_{H\ell}+C^{(3)}_{H\ell}),
                 \nonumber \\
V_{ZR1_l} &=& -\bar{g}_{Z1}  Q_e \hat{s}_W^2
                 -\hat{g}_Z Q_e \bar{s}_{W1}^2-\frac{\hat{g}_Z
                 \hat{v}^2}{2}(C_{He}),  \nonumber \\
V_{ZL1_u} &=& \bar{g}_{Z1} \left( I^3_u-Q_u \hat{s}_W^2 \right)
                 -\hat{g}_Z Q_u \bar{s}_{W1}^2-\frac{\hat{g}_Z
                 \hat{v}^2}{2}(C^{(1)}_{Hq}-C^{(3)}_{Hq}),
                 \nonumber \\
V_{ZR1_u} &=& -\bar{g}_{Z1}  Q_u \hat{s}_W^2
                 -\hat{g}_Z Q_u \bar{s}_{W1}^2-\frac{\hat{g}_Z
                 \hat{v}^2}{2}(C_{Hu}),  \nonumber \\
V_{ZL1_d} &=& \bar{g}_{Z1} \left( I^3_d-Q_d \hat{s}_W^2 \right)
                 -\hat{g}_Z Q_d \bar{s}_{W1}^2-\frac{\hat{g}_Z
                 \hat{v}^2}{2}(C^{(1)}_{Hq}+C^{(3)}_{Hq}),
                 \nonumber \\
V_{ZR1_d} &=& -\bar{g}_{Z1}  Q_d \hat{s}_W^2
                 -\hat{g}_Z Q_d \bar{s}_{W1}^2-\frac{\hat{g}_Z
                 \hat{v}^2}{2}(C_{Hd}).  
\end{eqnarray}  
These vertices are dependent on the following six Wilson
coefficient combinations:
\begin{equation}
C^{(1)}_{Hl}+C^{(1)}_{Hq}, C_{He}, C_{Hu}, C_{Hd},C^{(1)}_{Hq}, C^{(3)}_{Hq}.
\end{equation}  
There are seven Wilson coefficients total.  The combination $C^{(1)}_{H\ell}-C^{(3)}_{H\ell}$ cannot be probed in neutral current Drell-Yan. It is 
accessible in the charged-current Drell-Yan process.

At ${\cal O}(\kappa^2)$ there are three distinct contributions: the
${\cal O}(\kappa^2)$ corrections to the overall couplings, the ${\cal
  O}(\kappa)$ corrections to the ${\cal O}(\kappa)$ explicit vertex
factors, and the explicit ${\cal O}(\kappa^2)$ vertex factors.  The
explicit vertex corrections can be found in Ref.~\cite{Murphy:2020rsh}:
\begin{eqnarray}
V_{ZL2_l} &=& \bar{g}_{Z2} \left( I^3_e-Q_e \hat{s}_W^2 \right)
                 -\hat{g}_Z Q_e \bar{s}_{W2}^2-\bar{g}_{Z1} Q_e \bar{s}_{W1}^2
                 -\frac{\hat{g}_Z\bar{v}_1^2+\bar{g}_{Z1}\hat{v}^2}{2}(
                 C^{(1)}_{H\ell}+C^{(3)}_{H\ell})\nonumber \\
  &&-\frac{\hat{g}_Z \hat{v}^2}{4}(C^{(1)}_{\ell^2H^4D}+C^{(2)}_{\ell^2H^4D})
     \nonumber \\
V_{ZR2_l} &=& \bar{g}_{Z2} \left( -Q_e \hat{s}_W^2 \right)
                 -\hat{g}_Z Q_e \bar{s}_{W2}^2-\bar{g}_{Z1} Q_e \bar{s}_{W1}^2
                 -\frac{\hat{g}_Z\bar{v}_1^2+\bar{g}_{Z1}\hat{v}^2}{2}(
                 C_{He})\nonumber \\
             &&-\frac{\hat{g}_Z \hat{v}^2}{4}(C_{e^2H^4D}) \nonumber \\
V_{ZL2_u} &=& \bar{g}_{Z2} \left( I^3_u-Q_u \hat{s}_W^2 \right)
                 -\hat{g}_Z Q_u \bar{s}_{W2}^2-\bar{g}_{Z1} Q_u \bar{s}_{W1}^2
                 -\frac{\hat{g}_Z\bar{v}_1^2+\bar{g}_{Z1}\hat{v}^2}{2}(
                 C^{(1)}_{Hq}-C^{(3)}_{Hq})\nonumber \\
  &&-\frac{\hat{g}_Z \hat{v}^2}{4}(C^{(1)}_{q^2H^4D}-C^{(2)}_{q^2H^4D})
     \nonumber \\
V_{ZR2_u} &=& \bar{g}_{Z2} \left( -Q_u \hat{s}_W^2 \right)
                 -\hat{g}_Z Q_u \bar{s}_{W2}^2-\bar{g}_{Z1} Q_u \bar{s}_{W1}^2
                 -\frac{\hat{g}_Z\bar{v}_1^2+\bar{g}_{Z1}\hat{v}^2}{2}(
                 C_{Hu})\nonumber \\
             &&-\frac{\hat{g}_Z \hat{v}^2}{4}(C_{u^2H^4D}) \nonumber \\
V_{ZL2_d} &=& \bar{g}_{Z2} \left( I^3_d-Q_d \hat{s}_W^2 \right)
                 -\hat{g}_Z Q_d \bar{s}_{W2}^2-\bar{g}_{Z1} Q_d \bar{s}_{W1}^2
                 -\frac{\hat{g}_Z\bar{v}_1^2+\bar{g}_{Z1}\hat{v}^2}{2}(
                 C^{(1)}_{Hq}+C^{(3)}_{Hq})\nonumber \\
  &&-\frac{\hat{g}_Z \hat{v}^2}{4}(C^{(1)}_{q^2H^4D}+C^{(2)}_{q^2H^4D})
     \nonumber \\
V_{ZR2_d} &=& \bar{g}_{Z2} \left( -Q_d \hat{s}_W^2 \right)
                 -\hat{g}_Z Q_d \bar{s}_{W2}^2-\bar{g}_{Z1} Q_d \bar{s}_{W1}^2
                 -\frac{\hat{g}_Z\bar{v}_1^2+\bar{g}_{Z1}\hat{v}^2}{2}(
                 C_{Hd})\nonumber \\
             &&-\frac{\hat{g}_Z \hat{v}^2}{4}(C_{d^2H^4D}).
\end{eqnarray}  
A total of seven new dimension-8 coefficients enter the vertices at $\mathcal O(\kappa^2)$. These  coefficients introduce corrections to the 
cross section that scale as $\mathcal O(v^4/\Lambda^4)$, and are negligible for typical parameter choices.

We comment here briefly on the number of Wilson coefficients that enter our calculation. After accounting for the redefinitions of the input parameters, a total of 28 dimension-6 Wilson coefficients and 54 dimension-8 Wilson coefficients enter our result in the flavor-universal limit assumed here (we note that many enter only in  linear combinations, and cannot independently be probed). The number of contributing dimension-8 coefficients is reduced by the fact that since these couplings must interfere with the SM amplitude, the contributions from all scalar and tensor four-fermion operators, as well as all dipole operators, vanish in the massless fermion limit. This removes approximately 20 additional Wilson coefficients that would appear if fermion masses were not neglected.

\section{Renormalization group evolution of SMEFT coefficients}\label{RGE}

Most operators we consider 
are built out of quark vector and axial currents, which  
do not run in QCD \cite{Larin:1993tq}. This is the case for $C^{(1,3)}_{Hq}$, $C_{Hd}$, $C_{Hu}$ and $C_{Hud}$ in Eq. \eqref{eq:Z},
$C^{(1,3)}_{\ell q}$, $C^{}_{e u}$, $C^{}_{e d}$, $C^{}_{\ell u}$, $C^{}_{\ell d}$ and $C_{qe}$ in Eq. \eqref{eq:fourfermion}
and all the operators in Eq. \eqref{eq:qqllphi} and \eqref{eq:f2H4D}. The additional derivatives in the operators in Eq. \eqref{eq:fourfermion8} do not affect the renormalization of these operators under QCD.
The  operators $C^{(1,2,3,4)}_{q^2 H^2 D^3}$ 
$C^{(1,2)}_{u^2 H^2 D^3}$ and $C^{(1,2)}_{d^2 H^2 D^3}$
in Eq. \eqref{qqphiphid3}  have a covariant derivative acting on the quark field, which could in principle affect the renormalization of these operators. In the combinations that contribute to Drell-Yan, however, the covariant derivative can be moved on the weak bosons and Higgs fields, so that again  these operators do not renormalize in QCD. For example, we can write
\begin{eqnarray}
& &\frac{1}{2} \left(C^{(1)}_{u^2 H^2 D^3} - C^{(2)}_{u^2 H^2 D^3} \right)  i \bar u_R \gamma^\mu D^\nu u_R
\left\{ ( D_{(\mu} D_{\nu)} \varphi)^{\dagger } \varphi
- \varphi^{\dagger} ( D_{(\mu} D_{\nu)} \varphi) \right\} \nonumber \\
&&\propto\left(C^{(1)}_{u^2 H^2 D^3} - C^{(2)}_{u^2 H^2 D^3} \right) \, \bar u_R \gamma^\mu u_R\,  \partial^\nu ( \partial_{(\nu} Z_{\mu)}) + \ldots
\end{eqnarray}
where the $\ldots$ denote terms with more Higgs and weak gauge boson fields,
implying that $C^{(1)}_{u^2 H^2 D^3} - C^{(2)}_{u^2 H^2 D^3}$ does not run in QCD.

The coefficients of scalar operators in Eq.\ \eqref{eq:fourfermion}, $C_{\ell equ}^{(1)}$ and $C_{\ell edq}$, obey the same renormalization group equation as the quark masses.
The  dipole operators in Eq.\ \eqref{eq:dipole} and the tensor operator in Eq.\ \eqref{eq:fourfermion} have the same QCD anomalous dimension.    
The scalar and tensor operators satisfy
\begin{eqnarray}\label{rge1}
\frac{d}{d \log \mu}  C_S &=&  \frac{\alpha_s}{4\pi} \sum_n  \left( \frac{\alpha_s}{4\pi} \right)^n \gamma^{(n)}_S
C_S, \qquad  C_S \in \left\{C^{(1)}_{\ell e q u}, C^{}_{\ell ed q}\right\}  \nonumber \\
\frac{d}{d \log \mu}  C_T &=&  \frac{\alpha_s}{4\pi} \sum_n  \left( \frac{\alpha_s}{4\pi} \right)^n \gamma^{(n)}_T
C_T, \qquad  C_T \in \left\{C_{uW}, C_{uB}, C_{dW}, C_{dB}, C^{(3)}_{\ell e q u}\right\}, 
\end{eqnarray}
where the two loop anomalous dimensions are \cite{Misiak:1994zw,Vermaseren:1997fq,Degrassi:2005zd} 
\begin{eqnarray}\label{rge2}
\gamma_S^{(0)} &=& - 6 C_F, \qquad  \gamma_S^{(1)} = - \left( 3 C_F  + \frac{97}{3}  N_C - \frac{10}{3} n_f \right) C_F  ,\nonumber \\
\gamma_T^{(0)} &=& + 2 C_F, \qquad  \gamma_T^{(1)} = \left(\frac{257}{9}  N_C -19 C_F -\frac{26}{9} n_f\right) C_F .
\end{eqnarray}
Here $C_F = 4/3$, $N_C = 3$ and $n_f = 5$ is the number of light flavors.
The limits in Figure \ref{fig:bound2} are on scalar and tensor coefficients defined at the arbitrary scale $\mu_0 = 1$ TeV. They can be translated into limits at other scales by using Eq. \eqref{rge2}.

Finally, two-derivative operators in the same class as $C^{(2)}_{\ell^2 q^2 D^2}$ behave under QCD like twist-two operators. Their anomalous dimension is known to three loops  \cite{Gross:1974cs,Floratos:1977au,Gracey:2003mr}, and, at one loop:
\begin{equation}
 \frac{d}{d \log \mu} C^{(2)}_{\ell^2 q^2 D^2}  =  \frac{16}{3} \frac{\alpha_s C_F}{4\pi} C^{(2)}_{\ell^2 q^2 D^2}.
\end{equation}

\section{Cross sections}\label{CStables}

We report in this Appendix the contributions to the differential cross section $d\sigma/d m_{\ell \ell}$ from dimension-6 and dimension-8 SMEFT operators, at $\sqrt{S} = 8$ TeV and using the invariant mass binning of Ref. \cite{Aad:2016zzw}.
In Tables \ref{Vector6a} and \ref{Vector6b} we give results for the dimension-6 semi-leptonic vector operators  in Eq. \eqref{eq:fourfermion}, which interfere with the SM. For each Wilson coefficient, the first and second column report the values of the terms linear and quadratic in $C$, ($a^{(6)}$ and   $b^{(6)}$ in Eq. \eqref{sigmascheme}). 
Tables \ref{DipoleU}, \ref{DipoleD} and \ref{Scalar} give the cross sections induced by dipole, scalar and tensor operators. We do not show in this case the interference terms between $U(1)_Y$ and $SU(2)_L$ dipoles. The interference between scalar and tensor operators is negligible.
In Table \ref{Vector8} we show the cross section from dimension-8 derivative operators in Eq. \eqref{eq:fourfermion8}, while in Table 
\ref{ZFF8} the corrections from the operators in Eq. 
\eqref{qqphiphid3}. With the exception 
of $C^{(2)}_{q^2 e^2 H^2}$, the contributions from the operators in the class $\mathcal L_{\psi^4 H^2}$ can be obtained by rescaling the $a^{(6)}$ terms in Tables \ref{Vector6a} and \ref{Vector6b} according to  Eqs. \eqref{Vff1} and \eqref{Vff2}. We therefore only give the cross section 
induced by $C^{(2)}_{q^2 e^2 H^2}$, in the last column of 
Table \ref{Vector8}.

We do not give cross sections from the operators that only modify the $Z$ couplings, at dimension-6 in Eq. \eqref{eq:Z} and dimension-8 in Eq. \eqref{eq:f2H4D}.  

\begingroup
\squeezetable
\begin{table}
 \begin{tabular}{|c||cc| cc |c ||  }
 \hline
       & \multicolumn{2}{c|}{$C_{eu}$} & \multicolumn{2}{c|}{$C_{ed}$}  &  $C_{qe}$ \\
   bins & $a^{(6)}/\Lambda^2$ & $b^{(6)}/\Lambda^4$ 
    & $a^{(6)}/\Lambda^2$ & $b^{(6)}/\Lambda^4$ 
    & $a^{(6)}/\Lambda^2$ 
   \\
   \hline \hline
   116-130   & $-4.91(8)  \cdot 10^{-4}$ &  $2.25(5) \cdot 10^{-6}$  &                          
                $2.54(3)  \cdot 10^{-4}$ &  $2.32(4) \cdot 10^{-6}$  &
                $-2.59(7) \cdot 10^{-4}$   
                \\   
   130-150   & $-3.80(6)  \cdot 10^{-4}$ &  $2.45(5) \cdot 10^{-6}$  &
               $ 1.91(2)  \cdot 10^{-4}$ &  $2.45(3) \cdot 10^{-6}$  &
               $-1.90(3)  \cdot 10^{-4}$   
               \\
   150-175   & $-2.87(5)  \cdot 10^{-4}$ &  $2.61(5) \cdot 10^{-6}$  &
               $ 1.39(2)  \cdot 10^{-4}$ &  $2.52(4) \cdot 10^{-6}$  &
               $-1.40(2) \cdot 10^{-4}$                 \\
   175-200   & $-2.18(4)  \cdot 10^{-4}$ &  $2.73(5) \cdot 10^{-6}$  &
               $ 1.02(1)  \cdot 10^{-4}$ &  $2.55(3) \cdot 10^{-6}$  &
               $-1.03(1)  \cdot 10^{-4}$ 
               \\
   200-230   & $-1.67(3)  \cdot 10^{-4}$ &  $2.79(5) \cdot 10^{-6}$  &
               $ 7.54(11) \cdot 10^{-5}$ &  $2.53(3) \cdot 10^{-6}$  &
               $-7.77(10) \cdot 10^{-5}$ 
               \\
   230-260   & $-1.29(2)  \cdot 10^{-4}$ &  $2.84(6) \cdot 10^{-6}$  &
               $ 5.58(8)  \cdot 10^{-5}$ &  $2.48(4) \cdot 10^{-6}$  &
               $-5.85(9)  \cdot 10^{-5}$ 
               \\
   260-300   & $-9.79(21) \cdot 10^{-5}$ &  $2.84(6) \cdot 10^{-6}$  &
               $ 4.10(7) \cdot 10^{-5}$  &  $2.37(4) \cdot 10^{-6}$  &
               $-4.40(7) \cdot 10^{-5}$  
               \\
   300-380   & $-6.48(14) \cdot 10^{-5}$ &  $2.78(6) \cdot 10^{-6}$  &
               $ 2.58(5)  \cdot 10^{-5}$ &  $2.21(4) \cdot 10^{-6}$  &
               $-2.84(5)  \cdot 10^{-5}$ 
               \\
   380-500   & $-3.60(9)  \cdot 10^{-5}$ &  $2.57(6) \cdot 10^{-6}$  &
               $ 1.32(3)  \cdot 10^{-5}$ &  $1.89(4) \cdot 10^{-6}$  &
               $-1.53(3)  \cdot 10^{-6}$ 
               \\
   500-700   & $-1.61(4)  \cdot 10^{-5}$ &  $2.12(6) \cdot 10^{-6}$  &
               $ 5.52(13) \cdot 10^{-6}$ &  $1.45(4) \cdot 10^{-6}$ &
               $-6.72(18) \cdot 10^{-6}$   
               \\
   700-1000  & $-5.63(19) \cdot 10^{-6}$ &  $1.47(5) \cdot 10^{-6}$  &
               $ 1.78(6) \cdot 10^{-6}$ &  $9.28(29) \cdot 10^{-7}$  &
               $-2.30(7)  \cdot 10^{-6}$ 
               \\
   1000-1500 & $-1.41 (5)  \cdot 10^{-6}$ &  $7.54(27) \cdot 10^{-7}$ &
               $ 4.18 (17) \cdot 10^{-7}$ &  $4.40(19) \cdot 10^{-7}$ &
               $ -5.54(18) \cdot 10^{-7}$  \\
   \hline
 \hline
    & \multicolumn{2}{c|}{$C^{(1)}_{\ell q}$} & \multicolumn{2}{c|}{$C^{(3)}_{\ell q}$}  &  $C_{qe}$ \\
   bins & $a^{(6)}/\Lambda^2$ & $b^{(6)}/\Lambda^4$ & $a^{(6)}/\Lambda^2$ & $b^{(6)}/\Lambda^4$ &  $b^{(6)}/\Lambda^4$  \\
   \hline \hline
   116-130   & $-0.2(1.9)  \cdot 10^{-5}$  &  $4.54(6) \cdot 10^{-6}$  &                          
                $1.73(2)     \cdot 10^{-3}$  &  $4.54(6) \cdot 10^{-6}$  &
               $4.61(5) \cdot 10^{-6}$ 
                \\   
   130-150   & $-4.57(1.31)  \cdot 10^{-5}$ & $4.87(6) \cdot 10^{-6}$   &
               $ 1.21(1)  \cdot 10^{-3}$    & $4.87(6) \cdot 10^{-6}$  &
               $4.86(6) \cdot 10^{-6}$ 
               \\
   150-175   & $-5.99(91)  \cdot 10^{-5}$ &  $5.14(6) \cdot 10^{-6}$  &
               $ 8.44(10)  \cdot 10^{-4}$ &  $5.14(6) \cdot 10^{-6}$  &
               $5.12(6) \cdot 10^{-6}$
               \\
   175-200   & $-6.10(65)  \cdot 10^{-5}$ &  $5.27(7) \cdot 10^{-6}$  &
               $ 6.06(8)  \cdot 10^{-4}$ &  $5.27(7) \cdot 10^{-6}$  &
               $5.30(7) \cdot 10^{-6}$
               \\
   200-230   & $-5.59(45)  \cdot 10^{-5}$ &  $5.32(8) \cdot 10^{-6}$  &
               $ 4.46(6)  \cdot 10^{-4}$ &  $5.32(8) \cdot 10^{-6}$  &
               $5.34(7) \cdot 10^{-6}$
               \\
   230-260   & $-4.97(34)  \cdot 10^{-5}$ & $5.33(8) \cdot 10^{-6}$    &
               $ 3.33(5)  \cdot 10^{-4}$ &  $5.33(8) \cdot 10^{-6}$  &
               $5.31(8) \cdot 10^{-6}$\\
   260-300   & $-4.22(25) \cdot 10^{-5}$ &  $5.21(9) \cdot 10^{-6}$  &
               $ 2.46(4) \cdot 10^{-4}$  &  $5.21(9) \cdot 10^{-6}$  &
               $5.21(9) \cdot 10^{-6}$
               \\
   300-380   & $-3.15(16) \cdot 10^{-5}$ &  $4.98(10) \cdot 10^{-6}$  &
               $ 1.58(3)  \cdot 10^{-4}$ &  $4.98(10) \cdot 10^{-6}$  &
               $4.98(10) \cdot 10^{-6}$
                \\
   380-500   & $-2.01(10)  \cdot 10^{-5}$ &  $4.46(10) \cdot 10^{-6}$ &
               $ 8.39(19)  \cdot 10^{-5}$ & $4.46(10) \cdot 10^{-6}$  &
               $4.45(10) \cdot 10^{-6}$
               \\
   500-700   & $-1.00(4)  \cdot 10^{-5}$ &  $3.57(9) \cdot 10^{-6}$  &
               $ 3.66(10) \cdot 10^{-5}$ &   $3.57(9) \cdot 10^{-6}$ &
               $3.57(9) \cdot 10^{-6}$
               \\
   700-1000  & $-3.81(19) \cdot 10^{-6}$ &  $2.40(6) \cdot 10^{-6}$ &
               $ 1.23(4) \cdot 10^{-5}$ &  $2.40(7) \cdot 10^{-6}$  &
               $2.40(7) \cdot 10^{-6} $
               \\
   1000-1500 & $-1.04 (5)  \cdot 10^{-6}$ &  $1.19(4) \cdot 10^{-6}$ &
               $ 2.98 (10) \cdot 10^{-6}$ &  $1.19(4) \cdot 10^{-6}$ &
               $1.19(4) \cdot 10^{-6}$
               \\
   \hline \hline
    \end{tabular}
 \caption{Contributions to the differential cross section
 from vector-like four-fermion operators, in units of pb/GeV for the choice of  $\Lambda = 4$ TeV. For each coefficient, the first column denotes the interference with the SM, the second column the term quadratic in the SMEFT coefficient. The error denotes the PDF uncertainty. }\label{Vector6a}
\end{table}
\endgroup

\begingroup
\squeezetable
\begin{table}
\center
\begin{tabular}{||c||cc | cc ||  }
 \hline
          & \multicolumn{2}{c|}{$C_{\ell u}$} & \multicolumn{2}{c||}{$C_{\ell d}$}   \\
       bins         & $a^{(6)}/\Lambda^2$ & $b^{(6)}/\Lambda^4$ & $a^{(6)}/\Lambda^2$ & $b^{(6)}/\Lambda^4$  \\
   \hline \hline
   116-130   &  $-5.98(13)\cdot 10^{-5}$  &  $2.23(5) \cdot 10^{-6}$  
             &  $2.82(11) \cdot 10^{-5}$  &  $2.30(4) \cdot 10^{-6}$ \\   
   130-150   & $-9.42(19) \cdot 10^{-5}$  &  $2.43(5) \cdot 10^{-6}$  
             & $4.80(9) \cdot 10^{-5}$   &  $2.42(4) \cdot 10^{-6}$ \\
   150-175   & $-9.60(17) \cdot 10^{-5}$  &  $2.61(5) \cdot 10^{-6}$
             & $4.60(8) \cdot 10^{-5}$    &  $2.51(3) \cdot 10^{-6}$ \\
   175-200   & $-8.42(17) \cdot 10^{-5}$  &  $2.73(5) \cdot 10^{-6}$ 
             & $ 3.96(7) \cdot 10^{-5}$   &  $2.55(3) \cdot 10^{-6}$ \\
   200-230   & $-6.96(14) \cdot 10^{-5}$  &  $2.80(5) \cdot 10^{-6}$
             & $3.14(6) \cdot 10^{-5}$    &  $2.53(4) \cdot 10^{-6}$ \\
   230-260   & $-5.65(12) \cdot 10^{-5}$  &  $2.84(6) \cdot 10^{-6}$
             & $2.44(4) \cdot 10^{-5}$    &  $2.48(4) \cdot 10^{-6}$ \\
   260-300   & $-4.45(9) \cdot 10^{-5}$   &  $2.85(6) \cdot 10^{-6}$
             & $1.86(5) \cdot 10^{-5}$    &  $2.37(4) \cdot 10^{-6}$ \\
   300-380   & $-3.05(7)  \cdot 10^{-5}$  &  $2.78(6) \cdot 10^{-6}$ 
             & $1.21(2)   \cdot 10^{-5}$  &  $2.20(4) \cdot 10^{-6}$ \\
   380-500   & $-1.73(4)  \cdot 10^{-5}$ &  $2.56(6) \cdot 10^{-6}$  
             & $6.38(14) \cdot 10^{-6}$   &  $1.89(4) \cdot 10^{-6}$ \\
   500-700   & $-7.95(24) \cdot 10^{-6}$  &  $2.13(6) \cdot 10^{-6}$
             & $2.71(7) \cdot 10^{-6}$    &  $1.45(4) \cdot 10^{-6}$ \\
   700-1000  & $-2.77(7)  \cdot 10^{-6}$  &  $1.47(5) \cdot 10^{-6} $
             & $8.81(29) \cdot 10^{-7}$   &  $9.28(29) \cdot 10^{-7}$ 
   \\
   1000-1500 & $ -7.14(34) \cdot 10^{-7}$ & $7.55(27) \cdot 10^{-7}$ 
             & $2.06(10) \cdot 10^{-7}$   & $4.40(19) \cdot 10^{-7}$     \\
   \hline
 \end{tabular}
\caption{Contributions to the differential cross section
 from the operators $C_{\ell u}$ and $C_{\ell d}$, in units of pb/GeV for the choice of  $\Lambda = 4$ TeV. For each coefficient, the first column denotes the interference with the SM, the second column the term quadratic in the SMEFT coefficient. The error denotes the PDF uncertainty. }\label{Vector6b}
\end{table}
\endgroup

\begingroup
\squeezetable
\begin{table}
\center
\begin{tabular}{||c||cc  ||  }
\hline
      & $C^{}_{uW}$& $C^{}_{uB}$ \\
   bins   & $b^{(6)}_{}/\Lambda^4$ & $b^{(6)}_{}/\Lambda^4$ \\
   \hline \hline
   116-130   & $9.55(13) \cdot 10^{-6}$    &  $1.27(2)   \cdot 10^{-6}$  \\   
   130-150   & $5.15(8)  \cdot 10^{-6}$    &  $8.96(14)  \cdot 10^{-7}$  \\
   150-175   & $3.02(5)  \cdot 10^{-6}$    &  $6.39(10)  \cdot 10^{-7}$   \\
   175-200   & $1.97(3)  \cdot 10^{-6}$    &  $4.75(8)   \cdot 10^{-7}$   \\
   200-230   & $1.36(2)  \cdot 10^{-6}$    &  $3.54(6)   \cdot 10^{-7}$   \\
   230-260   & $9.76(18) \cdot 10^{-7}$    &  $2.68(5)   \cdot 10^{-7}$  \\
   260-300   & $7.01(13) \cdot 10^{-7}$    &  $1.99(4)   \cdot 10^{-7}$  \\
   300-380   & $4.38(9)  \cdot 10^{-7}$    &  $1.30(2)   \cdot 10^{-7}$  \\
   380-500   & $2.28(5)  \cdot 10^{-7}$    &  $6.96(16)  \cdot 10^{-8}$   \\
   500-700   & $9.71(26) \cdot 10^{-8}$    &  $3.01(8)   \cdot 10^{-8}$   \\
   700-1000  & $3.26(10) \cdot 10^{-8}$    &  $1.03(3)   \cdot 10^{-8}$   \\
   1000-1500 & $7.78(27) \cdot 10^{-9}$    &  $2.47(8)   \cdot 10^{-9}$   \\
   \hline
 \end{tabular}
\caption{Contributions to the differential cross section from the dipole operators $C_{uW}$ and $C_{uB}$, in units of pb/GeV for the choice of  $\Lambda = 4$ TeV.  The error denotes the PDF uncertainty. We do not show the interference term between $C_{uW}$ and $C_{uB}$. The Wilson coefficients are defined at the scale $\mu_0 = 1$ TeV. }\label{DipoleU}
\end{table}
\endgroup

\begingroup
\squeezetable
\begin{table}
\center
\begin{tabular}{||c||cc|c  c||  }
 \hline
            & $C^{}_{dW}$& $C^{}_{dB}$ & $C^{}_{eW}$& $C^{}_{eB}$    \\
    bins    & $b^{(6)}_{}/\Lambda^4$ & $b^{(6)}_{}/\Lambda^4$ &   $b^{(6)}_{}/\Lambda^4$ & $b^{(6)}_{}/\Lambda^4$  \\
   \hline \hline
   116-130   &  $9.54(11) \cdot 10^{-6}$  &  $1.27(1)  \cdot 10^{-6}$     
             &  $2.07(2) \cdot 10^{-5}$   &  $9.86(11) \cdot 10^{-7}$    \\   
   130-150   & $4.99(6) \cdot 10^{-6}$    &  $8.65(10) \cdot 10^{-7}$    
             & $1.06(1) \cdot 10^{-5}$    &  $5.40(7)  \cdot 10^{-7}$   \\
   150-175   & $2.82(3) \cdot 10^{-6}$    &  $5.98(7)  \cdot 10^{-7}$           
             & $5.89(7)\cdot 10^{-6}$     &  $3.27(4)  \cdot 10^{-7}$  \\
   175-200   & $1.78(2) \cdot 10^{-6}$    &  $4.29(6)  \cdot 10^{-7}$     
             & $3.69(5) \cdot 10^{-6}$    &  $2.28(4)  \cdot 10^{-7}$  \\
   200-230   & $1.17(2) \cdot 10^{-6}$    &  $3.05(4)  \cdot 10^{-7}$  
             & $2.48(3) \cdot 10^{-6}$    &  $1.62(3)  \cdot 10^{-7}$  \\
   230-260   & $8.14(12) \cdot 10^{-7}$   &  $2.23(3)  \cdot 10^{-7}$  
             & $1.73(3) \cdot 10^{-6}$    &  $1.21(2)  \cdot 10^{-7}$  \\
   260-300   & $5.52(10)\cdot 10^{-7}$    &  $1.58(3)  \cdot 10^{-7}$  
             & $1.21(2) \cdot 10^{-6}$    &  $8.75(17) \cdot 10^{-8}$  \\
   300-380   & $3.36(6) \cdot 10^{-7}$    &  $9.87(18) \cdot 10^{-8}$  
             & $7.45(14) \cdot 10^{-7}$   &  $5.76(12) \cdot 10^{-8}$ \\
   380-500   & $1.61(3) \cdot 10^{-7}$    &  $4.92(10) \cdot 10^{-8}$  
             & $3.75(8)\cdot 10^{-7}$     &  $3.01(6)  \cdot 10^{-8}$  \\
   500-700   & $6.38(16) \cdot 10^{-8}$   &  $1.99(5)  \cdot 10^{-8}$  
             & $1.56(4) \cdot 10^{-7}$    &  $1.31(4)  \cdot 10^{-8}$  \\
   700-1000  & $2.02(6) \cdot 10^{-8}$    &  $6.38(19)  \cdot 10^{-9}$  
             & $5.14(16)\cdot 10^{-8}$    &  $4.47(14)  \cdot 10^{-9}$  \\
   1000-1500 & $4.50(19)\cdot 10^{-9}$    &  $1.43(6)   \cdot 10^{-9}$  
             & $1.19(4) \cdot 10^{-8}$    &  $1.06(4)  \cdot 10^{-9}$   \\
   \hline \hline
   \end{tabular}
\caption{
Contributions to the differential cross section from the dipole operators $C_{dW}$, $C_{dB}$, $C_{eW}$, and $C_{eB}$, in units of pb/GeV for the choice of  $\Lambda = 4$ TeV.  The error denotes the PDF uncertainty. We do not show the interference term between $C_{dW}$ and $C_{dB}$
and $C_{eW}$ and $C_{eB}$. The Wilson coefficients are defined at the scale $\mu_0 = 1$ TeV.}\label{DipoleD}
\end{table}
\endgroup

\begingroup
\squeezetable
\begin{table}
\center
\begin{tabular}{||c||ccc ||  }
 \hline
            & $C_{\ell e d q}$ & $C^{(1) }_{\ell e q u}$ & $C^{(3) }_{\ell e q u}$     \\
    bins & $b^{(6)}_{}/\Lambda^4$ & $b^{(6)}_{}/\Lambda^4$ &   $b^{(6)}_{}/\Lambda^4$   \\
   \hline \hline
   116-130   & $5.42(8) \cdot 10^{-6}$    &  $5.28(10) \cdot 10^{-6}$   &  $1.61(2) \cdot 10^{-5}$  \\   
   130-150   & $5.53(7) \cdot 10^{-6}$    &  $5.53(10)  \cdot 10^{-6}$  &  $1.79(2) \cdot 10^{-5}$  \\
   150-175   & $5.54(7) \cdot 10^{-6}$    &  $5.76(11)  \cdot 10^{-6}$  &  $1.92(3) \cdot 10^{-5}$    \\
   175-200   & $5.46(8) \cdot 10^{-6}$    &  $5.92(12) \cdot 10^{-6}$   &  $2.00(3) \cdot 10^{-5}$   \\
   200-230   & $5.28(8) \cdot 10^{-6}$    &  $5.94(12) \cdot 10^{-6}$   & $2.04(3)  \cdot 10^{-5}$    \\
   230-260   & $5.07(9) \cdot 10^{-6}$    &  $5.89(12)  \cdot 10^{-6}$  & $2.05(4)  \cdot 10^{-5}$    \\
   260-300   & $4.78(8) \cdot 10^{-6}$    &  $5.77(12)  \cdot 10^{-6}$  & $2.04(4)   \cdot 10^{-5}$    \\
   300-380   & $4.29(8)\cdot 10^{-6}$     &  $5.49(12)  \cdot 10^{-6}$  & $1.97(4)   \cdot 10^{-5}$  \\
   380-500   & $3.55(8) \cdot 10^{-6}$    &  $4.88(12)  \cdot 10^{-6}$  & $1.81(4)   \cdot 10^{-5}$   \\
   500-700   & $2.60(7) \cdot 10^{-6}$    &  $3.86(11)  \cdot 10^{-6}$   & $1.52(4)  \cdot 10^{-5}$    \\
   700-1000  & $1.59(5)\cdot 10^{-6}$     &  $2.54(8)  \cdot 10^{-6}$   & $1.08(3)   \cdot 10^{-5}$   \\
   1000-1500 & $7.14(3) \cdot 10^{-7}$    &  $1.23(4)  \cdot 10^{-6}$   & $5.72(20)  \cdot 10^{-6}$   \\
   \hline
 \end{tabular}
\caption{
Contributions to the differential cross section from scalar and tensor four-fermion operators, in units of pb/GeV for the choice of  $\Lambda = 4$ TeV.  The error denotes the PDF uncertainty. 
We do not show the interference between the scalar and tensor operators
$C^{(1) }_{\ell e q u}$ and $C^{(3) }_{\ell e q u}$. The Wilson coefficients are defined at the scale $\mu_0 = 1$ TeV.
}\label{Scalar}
\end{table}
\endgroup

\begingroup
\squeezetable
\begin{table}
 \begin{tabular}{|c||c | c| c | c | c ||  }
\hline
       & $C^{(1)}_{e^2 u^2 D^2}$ &$C^{(1)}_{e^2 d^2 D^2}$  & $C^{(1)}_{e^2 q^2 D^2}$ & $C^{(1)}_{\ell^2 q^2 D^2}$ &  $C^{(3)}_{\ell^2 q^2 D^2}$ \\
   bins    & $a^{(8)}_{}/\Lambda^4$ & $a^{(8)}_{}/\Lambda^4$ &   $a^{(8)}_{}/\Lambda^4$   &   $a^{(8)}_{}/\Lambda^4$   &   $a^{(8)}_{}/\Lambda^4$  \\
   \hline \hline
   116-130   & $-4.64(7)  \cdot 10^{-7}$ &  $2.40(3)  \cdot 10^{-7}$  &                          
                $2.40(9) \cdot 10^{-7}$ &  $-0.7(1.8)\cdot 10^{-8}$  &
                $1.62(2) \cdot 10^{-6}$    
                \\   
   130-150   & $-4.64(8)  \cdot 10^{-7}$ &  $2.32(3) \cdot 10^{-7}$  &
               $ 2.38(7)  \cdot 10^{-7}$ &  $-5.5(1.6)        \cdot 10^{-8}$  &
               $ 1.47(2)  \cdot 10^{-6}$    
               \\
   150-175   & $-4.70(8)  \cdot 10^{-7}$ &  $2.27(3) \cdot 10^{-7}$  &
               $ 2.25(3)  \cdot 10^{-7}$ &  $-1.00(15)        \cdot 10^{-7}$  &
               $ 1.38(2) \cdot 10^{-6}$                  \\
   175-200   & $-4.75(9)  \cdot 10^{-7}$ &  $2.22(3) \cdot 10^{-7}$  &
               $ 2.24(4)  \cdot 10^{-7}$ &  $-1.35(14)        \cdot 10^{-7}$  &
               $ 1.33(2)  \cdot 10^{-6}$    
               \\
   200-230   & $-4.81(9)  \cdot 10^{-7}$ &  $2.16(3) \cdot 10^{-7}$  &
               $ 2.25(3) \cdot 10^{-7}$  &  $-1.61(13)        \cdot 10^{-7}$  &
               $ 1.28(2) \cdot 10^{-6}$   
               \\
   230-260   & $-4.80(10)  \cdot 10^{-7}$ &  $2.09(3) \cdot 10^{-7}$  &
               $ 2.16(4)  \cdot 10^{-7}$ &  $-1.86(13)        \cdot 10^{-7}$  &
               $ 1.24(2)  \cdot 10^{-6}$  
               \\
   260-300   & $-4.77(10) \cdot 10^{-7}$  &  $1.99(3) \cdot 10^{-7}$  &
               $ 2.15(4) \cdot 10^{-7}$  &  $-2.05(12)        \cdot 10^{-7}$  &
               $ 1.20(2) \cdot 10^{-6}$    
               \\
   300-380   & $-4.61(10) \cdot 10^{-7}$  &   $1.82(3) \cdot 10^{-7}$  &
               $ 2.02(4)  \cdot 10^{-7}$ &  $-2.26(12)        \cdot 10^{-7}$  &
               $ 1.12(2)  \cdot 10^{-6}$   \\ 
   380-500   & $-4.23(10)  \cdot 10^{-7}$ &  $1.56(3) \cdot 10^{-7}$  &
               $ 1.80(4)  \cdot 10^{-7}$ &  $-2.36(11)        \cdot 10^{-7}$  &
               $ 9.87(22)  \cdot 10^{-7}$   
               \\
   500-700   & $-3.48(10)  \cdot 10^{-7}$ &  $1.19(3) \cdot 10^{-7}$  &
               $ 1.45(4) \cdot 10^{-7}$  &  $-2.18(10)        \cdot 10^{-7}$ &
               $ 7.86(20) \cdot 10^{-7}$   \\
   700-1000  & $-2.41(8) \cdot 10^{-7}$ &   $7.60(24) \cdot 10^{-8}$  &
               $ 9.71(29) \cdot 10^{-8}$ &  $-1.65(8)         \cdot 10^{-7}$  &
               $ 5.27(16)  \cdot 10^{-7}$ \\
   1000-1500 & $-1.23 (4)  \cdot 10^{-7}$ &  $3.59(15) \cdot 10^{-8}$ &
               $ 4.88 (16) \cdot 10^{-8}$ &  $-9.17(51)         \cdot 10^{-8}$ &
               $ 2.62(9) \cdot 10^{-7}$  \\
   \hline
 \hline
       & $C^{(1)}_{\ell^2 u^2 D^2}$ & $C^{(1)}_{\ell^2 d^2 D^2}$ & $C^{(2)}_{q^2 e^2 H^2}$ &  & 
   \\
   bins    & $a^{(8)}_{}/\Lambda^4$ & $a^{(8)}_{}/\Lambda^4$ & $a^{(8)}_{}/\Lambda^4$ &  &\\
   \hline \hline
   116-130   & $-5.80(11) \cdot 10^{-8}$ &  $2.80(12) \cdot 10^{-8}$  & $-1.09(1)\cdot 10^{-6}$  & &  \\   
   130-150   & $-1.16(2)  \cdot 10^{-7}$ &  $5.84(9) \cdot 10^{-8}$  & $-5.43(6)\cdot 10^{-7}$ & &  \\
   150-175   & $-1.57(3)  \cdot 10^{-7}$ &  $7.64(10) \cdot 10^{-8}$  & $-2.61(4)\cdot 10^{-7}$ & &   \\
   175-200   & $-1.84(3)  \cdot 10^{-7}$ &  $8.61(13) \cdot 10^{-8}$  & $-1.31(2)\cdot 10^{-7}$  & &   \\
   200-230   & $-2.01(4)  \cdot 10^{-7}$ &  $9.05(13) \cdot 10^{-8}$  & $-6.64(14)\cdot 10^{-8}$ & &  \\
   230-260   & $-2.11(4)  \cdot 10^{-7}$ &  $9.15(14) \cdot 10^{-8}$  & $-3.37(10)\cdot 10^{-8}$ & &  \\
   260-300   & $-2.17(5)  \cdot 10^{-7}$ &  $9.05(16) \cdot 10^{-8}$  & $-1.56(7)\cdot 10^{-8}$ & & \\
   300-380   & $-2.16(5)  \cdot 10^{-7}$ &  $8.57(16) \cdot 10^{-8}$  & $-3.81(47)\cdot 10^{-9}$ & &  \\
   380-500   & $-2.04(5)  \cdot 10^{-7}$ &  $7.52(17) \cdot 10^{-8}$  & $ 1.23(26)\cdot 10^{-9}$ & &  \\
   500-700   & $-1.71(5)  \cdot 10^{-7}$ &  $5.83(15) \cdot 10^{-8}$  & $ 1.62(13)\cdot 10^{-9}$ & &  \\
   700-1000  & $-1.19(4)  \cdot 10^{-7}$ &  $3.76(12) \cdot 10^{-8}$  & $8.53(56)\cdot 10^{-10}$ & & \\
   1000-1500 & $-6.13(22) \cdot 10^{-8}$ &  $1.79(8)\cdot 10^{-8}$ & $2.60(19)\cdot 10^{-10}$& &   \\
   \hline \hline
    \end{tabular}
 \caption{Contributions to the differential cross section from 
 two-derivative dimension-8 operators, in units of pb/GeV for the choice of  $\Lambda = 4$ TeV. 
 The last column gives the cross section for the four-fermion two-Higgs operator $C^{(2)}_{q^2 e^2 H^2}$,  which cannot be obtained by rescaling the $a^{(6)}$ term in Table \ref{Vector6a}.
 }\label{Vector8}
\end{table}
\endgroup

\begingroup
\squeezetable
\begin{table}
\center
 \begin{tabular}{|c||c | c| c | c | c | c ||  }
 \hline
       & $C^{e}_{\ell^2 H^2 D^3}$ &$C^{(1)-(2)}_{q^2 H^2 D^3}$  & $C^{(3)-(4)}_{q^2 H^2 D^3}$  \\
   bins       & $a^{(8)}_{}/\Lambda^4$ & $a^{(8)}_{}/\Lambda^4$ & $a^{(8)}_{}/\Lambda^4$ \\
   \hline \hline
   116-130   & $7.79(7)  \cdot 10^{-7}$ &  $ 6.73(81)  \cdot 10^{-8}$  &                          
               $7.16(6)  \cdot 10^{-7}$    
                \\   
   130-150   & $ 4.11(4)  \cdot 10^{-7}$ &  $2.63(43) \cdot 10^{-8}$  &
               $ 3.66(4)  \cdot 10^{-7}$ 
               \\
   150-175   & $ 2.34(3)  \cdot 10^{-7}$ &  $1.19(22) \cdot 10^{-8}$  &
               $ 2.03(2)  \cdot 10^{-7}$ \\
   175-200   & $ 1.49(2)  \cdot 10^{-7}$ &  $4.2(1.3) \cdot 10^{-9}$  &
               $ 1.25(2) \cdot 10^{-7}$ 
               \\
   200-230   & $ 1.01(1)  \cdot 10^{-7}$ &  $6.8(8.2) \cdot 10^{-10}$  &
               $ 8.37(11) \cdot 10^{-8}$  
               \\
   230-260   & $ 7.10(11)  \cdot 10^{-8}$ &  $-2.9(5.9) \cdot 10^{-10}$  &
               $ 5.79(9)  \cdot 10^{-8}$ 
               \\
   260-300   & $ 4.97(8) \cdot 10^{-8}$  &  $-7.3(4.0) \cdot 10^{-10}$  &
               $ 4.08(7) \cdot 10^{-8}$  
               \\
   300-380   & $ 3.07(6) \cdot 10^{-8}$  &   $-1.46(23) \cdot 10^{-9}$  &
               $ 2.46(5)  \cdot 10^{-8}$  \\ 
   380-500   & $1.56(3)  \cdot 10^{-8} $ &  $-1.13(11) \cdot 10^{-9}$  &
               $ 1.26(3)  \cdot 10^{-8}$
               \\
   500-700   & $ 6.61(17)  \cdot 10^{-9}$ &  $-6.74(54) \cdot 10^{-10}$  &
               $ 5.28(13) \cdot 10^{-9}$     \\
   700-1000  & $ 2.21(6)  \cdot 10^{-9}$  &   $-2.97(22) \cdot 10^{-10}$  &
               $ 1.76(5) \cdot 10^{-9}$ \\
   1000-1500 & $5.21(17) \cdot 10^{-10}$ &  $-8.48(72) \cdot 10^{-11}$ &
               $ 4.14 (14) \cdot 10^{-10}$  \\
   \hline
 \hline 
    &  $C^{(1)-(2)}_{e^2 H^2 D^3}$ &  $C^{(1)-(2)}_{u^2 H^2 D^3}$ & $C^{(1)-(2)}_{d^2 H^2 D^3}$ \\
bins & $a^{(8)}_{}/\Lambda^4$ & $a^{(8)}_{}/\Lambda^4$ & $a^{(8)}_{}/\Lambda^4$ \\
   \hline \hline
   116-130   &  $-3.83(3)  \cdot 10^{-7}$  &
                $1.10(2) \cdot 10^{-7}$ &  $-5.72(9) \cdot 10^{-8}$    
                \\   
   130-150   &  $-1.73(2) \cdot 10^{-7}$  &
               $ 5.32(9) \cdot 10^{-8}$ &  $-2.70(5) \cdot 10^{-8}$     
               \\
   150-175   &  $-8.54(9)  \cdot 10^{-8}$  &
               $ 2.82(5) \cdot 10^{-8}$  &  $-1.38(3) \cdot 10^{-8}$   \\
   175-200    &  $-4.82(7)  \cdot 10^{-8}$  &
               $ 1.73(4)  \cdot 10^{-8}$ &  $-8.06(21) \cdot 10^{-9}$    
               \\
   200-230    &  $-2.97(4) \cdot 10^{-8}$  &
               $ 1.11(2) \cdot 10^{-8}$  &  $-5.11(12) \cdot 10^{-9}$ 
               \\
   230-260    &   $-1.94(3) \cdot 10^{-8}$  &
               $ 7.80(24)  \cdot 10^{-9}$  &  $-3.30(10) \cdot 10^{-9}$
               \\
   260-300    &  $-1.31(2) \cdot 10^{-8}$  &
               $ 5.54(16) \cdot 10^{-9}$  &  $-2.26(7) \cdot 10^{-9}$  
               \\
   300-380   &  $-7.59(13) \cdot 10^{-9}$  &
               $ 3.37(7)  \cdot 10^{-9}$ &  $-1.34(4) \cdot 10^{-9}$  \\ 
   380-500    &  $-3.59(9) \cdot 10^{-9}$  &
               $ 1.75(6)  \cdot 10^{-9}$ &  $-6.49(17) \cdot 10^{-10}$ 
               \\
   500-700    &  $-1.44(4)  \cdot 10^{-9}$ &
               $ 7.41(20) \cdot 10^{-10}$ &  $-2.54(7) \cdot 10^{-10}$   \\
   700-1000  &  $-4.68(14) \cdot 10^{-10}$  &
               $ 2.56(8)  \cdot 10^{-10}$ &  $-8.14(25) \cdot 10^{-11}$\\
   1000-1500  &  $-1.07(4) \cdot 10^{-10}$ &
               $ 6.13(21) \cdot 10^{-11}$ &  $-1.82(7) \cdot 10^{-11}$ \\
   \hline
 \hline
    \end{tabular}
 \caption{Contributions to the differential cross section from 
 momentum-dependent $Z$ couplings in Eq. \eqref{qqphiphid3}, in units of pb/GeV for the choice of  $\Lambda = 4$ TeV. }\label{ZFF8}
\end{table}
\endgroup

\bibliography{bibliography}

\end{document}